\documentclass[a4paper,fleqn,usenatbib]{mnras}
\usepackage{newtxtext,newtxmath}

\usepackage[T1]{fontenc}
\usepackage{ae,aecompl}
\usepackage{graphicx}   
\usepackage{amsmath}    
\usepackage[normalem]{ulem}
\usepackage[usenames]{color}

\newif\ifredline
\definecolor{DarkGreen}{rgb}{0.64,0.80,0.35}

\redlinefalse 

\ifredline
\newcommand\corr[1]{{\color{red} #1}} 
\else
\newcommand\corr[1]{#1} 
\renewcommand{\sout}[1]{}
\fi

\newcommand{\changaMM}{{MANGA}}
\newcommand{\be}{\begin{eqnarray}}
\newcommand{\ee}{\end{eqnarray}}

\newcommand{\grad}{\ensuremath{\boldsymbol{\nabla}}}
\newcommand{\vel}{\ensuremath{\boldsymbol{v}}}

\newcommand{\ddt}[1]{\ensuremath{\frac{\partial #1}{\partial t}}}
\newcommand{\state}{\ensuremath{\boldsymbol{\mathcal{U}}}}
\newcommand{\charge}{\ensuremath{\boldsymbol{U}}}
\newcommand{\acharge}{\ensuremath{\boldsymbol{\alpha}}}
\newcommand{\psicharge}{\ensuremath{\boldsymbol{\psi}}}

\newcommand{\flux}{\ensuremath{\boldsymbol{\mathcal{F}}}}
\newcommand{\fluxV}{\ensuremath{\boldsymbol{F}}}
\newcommand{\source}{\ensuremath{\boldsymbol{\mathcal{S}}}}
\newcommand{\sourceV}{\ensuremath{\boldsymbol{S}}}

\newcommand{\normal}{\ensuremath{\hat{\boldsymbol{n}}}}
\newcommand{\pt}{\ensuremath{\boldsymbol{p}}}
\newcommand{\nb}{\ensuremath{\boldsymbol{n}}}
\newcommand{\meshv}{\ensuremath{\boldsymbol{w}}}
\newcommand{\facev}{\ensuremath{\boldsymbol{\tilde{w}}_{ij}}}
\newcommand{\facer}{\ensuremath{\boldsymbol{\tilde{r}}_{ij}}}
\newcommand{\meshr}{\ensuremath{\boldsymbol{r}}}
\newcommand{\cmr}{\ensuremath{\boldsymbol{c}}}

\title{A Moving Mesh Hydrodynamic Solver for ChaNGa}

\author[Chang, Wadsley, \& Quinn]{Philip Chang$^1$\thanks{E-mail: chang65@uwm.edu}, James Wadsley$^2$, and Thomas R. Quinn$^3$\\
$^1$ Department of Physics, University of Wisconsin-Milwaukee, 3135 North Maryland Avenue, Milwaukee, WI 53211, USA\\
$^2$  Department of Physics \& Astronomy, McMaster University,
Hamilton, ON, L8S 4M1, Canada\\
$^3$  Department of Astronomy, Box 351580, University of Washington,
Seattle, WA 98195-1580, USA}

\begin{document}
\label{firstpage}
\pagerange{\pageref{firstpage}--\pageref{lastpage}}
\maketitle

\begin{abstract}
We describe the structure and implementation of a moving-mesh hydrodynamics solver in the large-scale parallel code, Charm N-body GrAvity solver (ChaNGa).  While largely based on the algorithm described by Springel (2010) \corr{\sout{and}that is} implemented in AREPO, our algorithm differs a few aspects.  We describe our use of the Voronoi tessellation library, VORO++, to compute the Voronoi tessellation directly.  We also incorporate some recent advances in gradient estimation and reconstruction that gives better accuracy in hydrodynamic solutions at minimal computational cost. We validate this module with a small battery of test problems against the smooth particle hydrodynamics solver included in ChaNGa.  Finally, we study one example of a scientific problem involving the mergers of two main sequence stars and highlight the small quantitative differences between smooth particle and moving-mesh hydrodynamics. We close with a discussion of anticipated future improvements and advancements.
\end{abstract}

\section{Introduction}

Numerical simulations have played a crucial role in understanding the hydrodynamics (HD) and magnetohydrodynamics (MHD) of gas, stars, disks, galaxies, large-scale structure, and other astrophysical phenomenon.  In this regard, two dominant methodologies have emerged to numerically solve the HD and MHD equations: smooth particle hydrodynamics (SPH) and grid-based solvers.

SPH is based upon the Lagrangian view of the Euler equations where the sampling of a fluid is determined from a finite number of particles, and fluid quantities like density and pressure are determined by computing a smoothing kernel over a number of neighbors.  The Lagrangian nature of SPH allows it to conserve linear and angular momentum, but comes at the expense of comparatively poor resolution of shocks due to its smoothing nature.  On the other hand, grid based methods have superior shock capturing abilities due to the use of Godonov schemes, but suffers from grid effects and possible violations of Galilean invariance.

Arbitrary Lagrangian-Eulerian (ALE) schemes have been devised as an effort to capture the best characteristics of both approaches (for a review see \citealt{ALEBook}). \citet{Noh1964} proposed the first 2-D ALE scheme where the initial grid was continuously deformed as the solution progressed.
In 3-d, \citet{1995ApJS...97..231G}, \citet{1998ApJS..115...19P} also proposed schemes based on similar ideas.  These moving-mesh (MM) methods suffered from mesh distortion or mesh tangling.  To correct this, these methods remapped the fluid onto a new mesh, which is inherently a diffusive operation.  As a result, these methods were not widely adopted in astrophysics with the exception of supernova modeling \citep{2008ApJS..179..209M}.

The need to avoid grid tangling and the expensive and diffusive grid remapping operations lead to schemes where the remapping happens continuously or in localized patches.  \citet{1987JCoPh..70..397B} proposed constructing a Voronoi diagram on a moving field of fluid markers, and then using this diagram to construct finite difference operators to solve the fluid equations.  Later, \citet{1995MNRAS.277..655W} proposed a method by which a Delaunay tessellation was constructed on a field of moving fluid elements on which finite volume methods are used to solve the fluid equations.  These two methods both avoid the effect of grid tangling and diffusive global grid remapping, but they were also only limited to 2-D, and were, perhaps, ahead of their time.

\citet[][hereafter S10]{2010MNRAS.401..791S} described a usable ALE scheme that has proven successful.  Implemented into the code, AREPO, the scheme relies on a Voronoi tessellation to generate well-defined and unique meshes for an arbitrary distribution of points. AREPO was built on top of the Gadget codebase \citep{2005MNRAS.364.1105S}, a widely used massively parallel Tree-SPH code, and has been widely applied to a number of problems.
The use of Voronoi tessellations allows a unique mesh to be defined, and these meshes deform continuously under the movement of the mesh generating points.  These cells have well defined volumes and faces on which finite volume methods can be applied.  The implementation of AREPO detailed in S10 has been extended to include magnetic fields \citep{2011MNRAS.418.1392P,2014MNRAS.442...43M,2016MNRAS.463..477M}, better convergence \citep{2016MNRAS.455.1134P,2015MNRAS.452.3853M}, and new physics, such as cosmic rays \citep{2016MNRAS.462.2603P,2017MNRAS.465.4500P} and have been used in a number of different problems including cosmological galaxy formation \citep[see for instance][]{2014MNRAS.444.1518V}, and stellar mergers \citep{2015ApJ...806L...1Z,2016ApJ...816L...9O}.

The scheme proposed in S10 has led to a number of ALE codes including TESS\citep{2011ApJS..197...15D}, FVMHD3D\citep{2012ApJ...758..103G}, ShadowFax\citep{2016A&C....16..109V}, RICH\citep{2015ApJS..216...35Y}, and DISCO\citep{2016ApJS..226....2D}.  In addition, the general scheme of determining the geometry from an arbitrary collection of points has also led to derivative methods such as GIZMO \citep{2015MNRAS.450...53H}.

Both AREPO and GIZMO are built on top of the Gadget codebase \citep{2005MNRAS.364.1105S}, which demonstrates that SPH codes can be modified to an ALE scheme. A similar code, Gasoline \citep{2004NewA....9..137W}, has been developed by James Wadsley, Joachim Stadel and Thomas Quinn and used in a number of different problems.  A successor to Gasoline that has been under heavy development over the last decade is called ChaNGa \citep{Jetley2008,Jetley2010,2015ComAC...2....1M}.  In addition to SPH, ChaNGa includes standard physics modules that have been ported from Gasoline including
metal line cooling, star formation, turbulent diffusion of metals and
thermal energy, and supernovae feedback \citep{2006MNRAS.373.1074S,2010MNRAS.407.1581S}.  The usefulness of ChaNGa in galaxy simulations has been demonstrated in the AGORA project \citep{2014ApJS..210...14K,2016ApJ...833..202K}.  The most up-to-date description of the algorithms in Gasoline and ChaNGa is given in \citet{Wadsley2017}

ChaNGa is also unique among astrophysical codes\footnote{However, a Charm++ version of ENZO is currently under development by James Bordner and Michael Norman \corr{and FVMHD3D by \citet{2012ApJ...758..103G} also uses the Charm++ language}.} in that it uses the Charm++ language and run-time system \citep{KaleKrishnan96} for parallelization rather than a custom message-passing interface design.  The use of Charm++ promises that ChaNGa will be much more scalable compared to previous astrophysical codes.  The message driven paradigm of Charm++ allows ChaNGa to be latency tolerant and overlap communication (significant in the SPH calculation) with computation (significant in the gravity calculation).  The concept of over-decomposition allows the work to be composed into objects independently of the number of processors, and the Charm++ adaptive run-time system distributes these objects among the physical processors in order to balance the computational load and minimize communication.  Using these features, ChaNGa has demonstrated strong scaling on single timestepping problems with $12$ billion particles to 512K cores (with 93\% efficiency) and on multi-timestepping problems with 52 million particles to 128K cores on Blue Waters\citep{2015ComAC...2....1M}.  In the era of exascale computing, such scalability becomes increasingly important.

We implement a MM hydrodynamic solver based on the algorithm describe in S10 in the latest public release of ChaNGa (version 3.2).  This module does not have an official name, but for the purposes of this paper, we will call it \changaMM\ rather than calling it ``the MM hydrodynamics solver module for ChaNGa.''  This is mainly for reasons of convenience in nomenclature.  Our implemention of \changaMM\ differs from S10 in several ways.  First, rather than generating the Delaunay tessellation and computing its dual to produce the Voronoi tessellation, we compute the Voronoi tessellation \corr{\textit{directly}} using the publicly available library, VORO++ \citep{2009Chaos..19d1111R}.  Second, we use an improved gradient estimate from \citet{2016MNRAS.459.1596S} and (optionally) use a total variation diminishing (TVD) limiter proposed by \citet{2011ApJS..197...15D}.  Finally, we perform the reconstruction of face centered values from the half-time step state vector and gradient estimates.

This paper is organized as follows.  We outline the basic algorithm in \S~\ref{sec:outline}.  We then describe the steps of the algorithm in brief, but in sufficient detail to allow this paper to serve as a reference for future developers and users of \changaMM.  In particular, we detail the differences between \changaMM\ and the implementation of AREPO described in S10.  We describe the construction of the Voronoi tessellation in \S~\ref{sec:voronoi} using the VORO++ library.  We then describe the hydrodynamic algorithm in \S~\ref{sec:hydro} with particular emphasis on the reconstruction of face-centered quantities in \S~\ref{sec:reconstruction}, the determination of mesh-generating point velocities and timesteps in \S~\ref{sec:centering}, the use of the entropy versus energy evolution equations in \S~\ref{sec:entropy}, and initialization in \S~\ref{sec:initialization}.  We then show the performance on \changaMM\ compared to the SPH solver in ChaNGa on a number of 3-D test problems in \S~\ref{sec:test problems} including the Sod shock-tube problem, the Sedov-Taylor point explosion, the Gresho-Chan vortex, the Evrard collapse problem, a stellar hydrostatic balance problem, and a stellar merger.  We describe a number of improvements targeted for the future in \S~\ref{sec:future} and close with some conclusions in \S~\ref{sec:conclusions}.

\section{Outline of the Algorithm}\label{sec:outline}

The ALE algorithm that is implemented in \changaMM\ is summarized as follows. Much of the algorithm follows that of S10, but with notable differences.
\begin{enumerate}
 \item We determine a valid Voronoi tessellation of the mesh-generating points using the VORO++ library.
 \item Using the volume of the Voronoi cell and integral quantities, $\charge$, the conserved and primitive variables are determined.  The local gradients and half-time-step conserved variables are calculated.
 \item The half-time-step gradients are calculated and used along with the half-time-step conserved variables to reconstruct the half-time-step face-centered quantities. An appropriate (optionally total variation diminishing) gradient limiter is applied.
 \item An Harten-Lax-van Leer-Contact (HLLC) or Harten-Lax-van Leer (HLL) approximate Riemann solver estimates the numerical flux in the rest frame of the moving faces of the half-time-step state.  This flux is transformed back to the ``lab'' frame to determine the changes to the state.
 \item The new state $\charge'$ is determined from the fluxes. New velocities for the mesh generating points are determined. The mesh generating points drift by a full time step.
\end{enumerate}

We refer the reader to S10 for an excellent and detailed discussion.  Here we will briefly describe our methodology and highlight differences between our implementation and that of S10.

\section{Direct Construction of the Voronoi Tessellation} \label{sec:voronoi}

The key insight of S10 to allow the construction of a robust ALE scheme was the use of a Voronoi tessellation to generate a mesh from an arbitrary collection of points on which the (M)HD equations can be solved with a finite-volume scheme.
The Voronoi tessellation has some important properties that make it particularly amendable for the construction of an ALE scheme.  First, the tessellation is unique.  Second, the tessellation varies smoothly and continuously under the motion of the underlying mesh generation points.  By smoothly and continuously, we mean that the neighbors do not suddenly change due to perturbation in the position of the mesh generating points, and the quantities that define the cell, i.e., the volume \corr{ of the cell\sout{, number of faces,}} and areas of the faces change smoothly under a perturbation of the mesh generating points.  This allows neighbors to be well defined and change smoothly across a time-step.

A Voronoi tessellation can be created in two ways.
First, it can be created by determining a Delaunay triangulation, which in 3-D is a partition of a space by tetrahedra where the vertices of the tetrahedra are given by the mesh-generating points, and no points are contained in the circumsphere of any given tetrahedra.  The Voronoi tessellation is then computed by computing the dual to the Delaunay tessellation.  Second, it can be computed directly.

The first approach is taken by S10, who describes the generation and parallelization of the Delaunay tessellation from which the Voronoi tessellation can be computed.  This algorithm is implemented in AREPO, TESS, FVM3D, and RICH because the quality of S10's description. Here, the empty circumsphere property uniquely determines the tetrahedra partition of the space.  The Voronoi tessellation is then computed by determining the centers of the circumspheres for each tetrahedra.  These centers are the vertices that make up the Voronoi tessellation.

The second case is the approach described in this paper. Using the publicly available Voronoi tessellation library Voro++\footnote{\url{http://math.lbl.gov/voro++/}}\citep{2009Chaos..19d1111R}, we compute the tessellation directly. We first enclose a point \pt\ about which we wish to compute its Voronoi cell with a large rectangular cell, which is the starting guess for the Voronoi cell.  The geometry of this large cell is irrelevant, but it must be much larger than the Voronoi cell that will be computed.  Using a rapid nearest-neighbor search algorithm, we can search for all neighbors \nb\ up to a radius $r_s$, which we will presume is arbitrary.  We order the neighbors, \nb, by distance, and starting from the nearest \nb, we compute the plane that bisects the line connecting \pt\ and \nb.  If the plane does not partition the Voronoi cell, i.e., intersect any of the faces of the Voronoi cell, then \nb\ does not share a common face with \pt.  If it does, then \nb\ shares a face with \pt, and the partition of the Voronoi cell by this plane forms a new face of the cell, whose vertices we compute.  We then update the Voronoi cell with the new face and the modifications to existing faces.  We continue bisecting the Voronoi cell, with \nb's of increasing distance from \pt\ until the distance between \pt\ and \nb\ is more than twice that between \pt\ and the most distance vertex in the Voronoi cell.  All points greater than this distance can no longer partition the Voronoi cell, which is now complete.

We have tested the Voro++ library and found that it has very high performance.  Using its native containers, Voro++ is able to compute 100K 3-D Voronoi cells per second on a single core of an Intel core i7-3770 running at 3.40GHz.  Moreover, because the calculations are entirely local in the neighborhood of each point, it is straightforward to parallelize provided that a sufficiently fast neighbor search exists.  We have found that the native Voro++ containers are expensive to instantiate, so we compute the Voronoi tessellation on a cell-by-cell basis.  This reduces the performance of the Voro++ library substantially, but we are still able to generate 10K cell per second per core, which is adequate performance.  Part of this poorer performance is due to the crude criterion to test for completion of a Voronoi cell that we use, i.e., all remaining neighbors are at least twice the distance to the \corr{\sout{furthest}farthest} vertex.  The native containers in Voro++ perform additional tests, where points only on one side of the cells is tested.  S10 also described more optimized search strategies.  The other reason for this slowdown is that the domain decomposition is totally arbitrary and not optimized for the construction of Voronoi tessellation.  Optimization of the construction of Voronoi cells is an improvement targeted for \changaMM, but its performance is sufficient for now.


\section{Hydrodynamics on a Moving Voronoi Tessellation}\label{sec:hydro}

\changaMM\ solves the Euler equations and evolution equation for entropy, which written in conservative form is:
\be
\ddt{\rho} + \grad\cdot\rho\vel &=& 0 \label{eq:continuity}\\
\ddt{\rho\vel} + \grad\cdot\rho\vel\vel + \grad P &=&-\rho\grad\Phi\label{eq:momentum}\\
\ddt{\rho e} + \grad\cdot\left(\rho e + P\right)\vel &=& -\rho\vel\cdot\grad\Phi\label{eq:energy}
\ee
where $\rho$ is the density, $\vel$ is the velocity, $\Phi$ is the gravitational potential, $e= \epsilon + v^2/2$ is the specific energy, $\epsilon$ is the internal energy, and $P(\rho, \epsilon)$ is the pressure.  Equations (\ref{eq:continuity}) - (\ref{eq:energy}) can be written in a compact form by introducing a state vector $\state=(\rho, \rho\vel, \rho e)$:
\be
\ddt{\state} + \int \grad\cdot\flux dV = \source\label{eq:state}
\ee
where $\flux=(\rho\vel, \rho\vel\vel, (\rho e + P)\vel)$ is the flux function, and $\source = (0, -\rho\grad\Phi, -\rho\vel\cdot\grad\Phi$) is the source function.

To solve equation (\ref{eq:state}), we adopt the same finite volume strategy as in S10.  We refer the interested reader to S10 for a more detailed discussion of the scheme.  Here, we will only briefly describe the scheme to document the algorithm we have implemented and to highlight the differences between our scheme and that of S10.

For each cell, the integral over the volume of the $i$th cell defines the charge of the $i$th cell, $\charge_i$, to be
\be
\charge_i = \int_i \state dV = \state_i V_i,
\ee
where $V_i$ is the volume of the cell.
As in S10, we then use Gauss' theorem to convert the volume integral over the divergence of the flux in equation (\ref{eq:state}) to a surface integral:
\be
\int_i \grad\cdot\flux dV = \int_i \flux\cdot\normal dA
\ee
We now take advantage of the fact that the volumes are Voronoi cells with a finite number of neighbors to define a integrated flux
\be
\sum_{j \in {\rm neighbors}} \fluxV_{ij} A_{ij} = \int_i \flux\cdot\normal dA,
\ee
where $\fluxV_{ij}$ and $A_{ij}$ are the average flux and area of the common face between cells $i$ and $j$.
The discrete time evolution of the charges in the system is given by:
\be
\charge_i^{n+1} = \charge_i^n + \Delta t \sum_j \hat{\fluxV}_{ij} A_{ij} + \Delta t\sourceV_i, \label{eq:time evolution}
\ee
where $\hat{\fluxV}_{ij}$ is an estimate of the half-time-step flux between the initial, $\charge_i^n$, and final states $\charge^{n+1}_i$ and is discussed in \S~\ref{sec:reconstruction}, and $\sourceV_i^{(n+1/2)} = \int_i \source dV$ is the time-averaged integrated source function.

We estimate the flux across each face, $\hat{\fluxV}_{ij}$, using an approximate Riemann solver.
As Riemann solvers for irregular cells in multidimensions do not exist, we follow the prescription of S10. We compute the 1-D fluxes across each face in the rest frame of that face and then collectively apply them per time-step.  The steps involved are:
\begin{enumerate}
 \item Estimate the velocity $\facev$ of the face -- following S10, the face velocities are:
 \be
 \facev = \frac{(\meshv_i - \meshv_j)\cdot(\facer - 0.5(\meshr_j+\meshr_i))}{|\meshr_j - \meshr_i|}\frac{\meshr_j - \meshr_i}{|\meshr_j - \meshr_i|} + \bar{\boldsymbol{w}}_{ij},
 \ee
 where $\bar{\boldsymbol{w}}_{ij} = 0.5(\meshv_i + \meshv_j)$ is the average velocity of the two mesh generating points and \facer\ is the face center between cells i and j.
\item Estimate the half-time-step state vector (in the rest frame of the moving face) at the face center (\facer) between the neighboring $i$ and $j$ cells via linear reconstruction -- we discuss this reconstruction in \S~\ref{sec:reconstruction}.
 \item Boost the state vector from the ``lab'' frame to the rest frame of the face center and rotate the state vector such that the x-axis points along the outward normal of the face, i.e., in the direction from $i$ to $j$.
 \item Estimate the flux using an HLL or HLLC Riemann solver implemented following \citet{toro2009riemann}. Here we found that both Riemann solvers give acceptable performance, though the HLL solver is more diffusive for problems that involve large gradients integrated over long timescale, i.e., hydrostatic balance.  By default, we choose HLLC.
 \item Boost the solved flux back into the ``lab'' frame.
\end{enumerate}
We can then use the estimated fluxes to time evolve the charges, $\charge_i$, following equation (\ref{eq:time evolution}).

The inclusion of (self-)gravity involves incorporating the gravitational potential into the momentum and energy equations.  Here our formalism takes care of the momentum equation modulo a proper definition for the time average integrated source function.  Here we follow the suggestion of S10 and define
\be
\int_i \rho_i \grad\Phi_i dV = \frac {1}{2}\left(m^{(n)}_i\grad\Phi_i^{(n)} + m^{(n+1)}_i\grad\Phi_i^{(n+1)} \right),
\ee
where $m = \int_i \rho dV$ is the integrated mass per cell. The source for the energy equation can similarly be defined by S10 as the ``standard approach'':
\be\label{eq:standard approach}
\int_i \rho_i \grad\Phi_i\cdot\vel dV = \frac {1}{2} m^{(n)}_iv^{(n)}_i\grad\Phi_i^{(n)} + \nonumber \\  \frac 1 2 m^{(n+1)}_iv^{(n+1)}_i\grad\Phi_i^{(n+1)},
\ee
but as S10 noted, a better scheme is to utilize the Green-Gauss theorem to produce an ``improved approach.'' As described by S10, it involves expanding the velocity into the cell velocity component and the fluid velocity component (relative to the cell) in the above integral.  The component that involves the cell velocity can be integrated over volume and time to give the change in the gravitational potential energy due to the movement of the entire cell.  The component that involves the relative (to the mesh generating point) fluid velocity can be converted to a surface integral and integrated over time to give the change in the gravitational potential energy of the cell due to advection of gravitational potential energy across the cell faces.  The relevant equation for the ``improved approach'' is
\be
\int_i \rho_i \grad\Phi_i\cdot\vel dV  = m_i\grad\Phi_i \cdot\meshv_i + \nonumber \\
\int_i\rho\left[\left(\vel - \meshv_i\right)\cdot\normal\right]\left(\boldsymbol{r} - \meshr_i\right)\cdot\grad\Phi_i  dA.
\ee
Here as in S10, the density in the first term on the RHS can be integrated over the volume of the cell to give the total mass of the cell, holding the gravitational potential and cell velocity to be constant.  The second term can be approximated noting that term, $\int_i\rho\left[\left(\vel - \meshv_i\right)\cdot\normal\right]dV$, is the mass flux across all the faces and the term $\left(\boldsymbol{r} - \meshr_i\right)\cdot\grad\Phi_i $ is the change in potential going from the face to the cell center. We approximate this second term by using the mass flux (across each face) calculated by the Riemann solver and setting $\left(\boldsymbol{r} - \meshr_i\right)\cdot\grad\Phi_i = \left(\meshr_j - \meshr_i\right)\cdot\grad\Phi_i/2$.   Both the ``standard'' and ``improved'' approaches are implemented in \changaMM, but the ``improved'' approach is used by default.

The gradient of the gravitational potential, $\grad\Phi$, i.e., the gravitational acceleration, can either be user-specified, derived from the solution of Poisson's equation:
\be
 \nabla^2\Phi = 4\pi G\rho
\ee
for the case of self-gravity, or a hybrid of  the two.  For self-gravity, we use the tree-based solver in ChaNGa where each tree node contains the multipole mass moments up to hexadecapole order.  The tree traversal algorithm for gravity is based on PKDGrav as described in \citet{2001PhDT........21S}.  It is similar to the Barnes-Hut \citep{1986Natur.324..446B} algorithm with some optimizations; for a few more details see \citet{2015ComAC...2....1M}.
Each leaf node of the tree contains a number (by default, 12 or less) of mesh generating points or other types of particles.  The mass distribution, $\rho$, is represented by a cubic spline softened density distribution with fixed softening centered at each mesh generation point.  This is consistent with how ChaNGa represents the density distribution from collisionless particles.

\subsection{Gradient Estimation and Reconstruction of Conserved Quantities}\label{sec:reconstruction}

A second order accurate code in time and space demands an appropriate estimate for the state vector at the face centers of each cell at the half timestep. In \changaMM, we first make an estimate for the state vector at the half timestep, use this half timestep state vector to compute the spatial gradient, and then use the spatial gradient to estimate the state vector at the face centers.  For our estimate of the state vector at the half timestep, we use an estimate for the time derivative from equation (\ref{eq:state})
\be
\ddt{\state}^{(n)} = -\grad\cdot\flux^{(n)} + \source^{(n)},\label{eq:time derivative estimate}
\ee
where we express $\grad\cdot\flux^{(n)}$ in terms of gradients of the state vector: $\grad{\state}_i^{(n)}$.
The half-time step state vector can then be determined:
\be
\state_i^{(n+1/2)} = \state_i^{(n)} + \ddt{\state_i^{(n)}} \frac {\Delta t} 2 .
\ee
The state vector on face of the $i$th and $j$th cell on the ($n+1/2$)-th timestep, $\tilde{\state}_{ij}^{(n+1/2)}$, is then
\be
\tilde{\state}_{ij}^{(n+1/2)} = \state_i^{(n+1/2)}  + (\facer - \cmr_i)\cdot\grad\state_i^{(n+1/2)}, \label{eq:state vector face}
\ee
where $\cmr_i$ is the center of mass of the $i$th cell.

Equations (\ref{eq:time derivative estimate}) and (\ref{eq:state vector face}) require an estimate for the gradient of the state vector.  Here, we follow the procedure of \citet{2016MNRAS.459.1596S} who improved upon the prescription of S10 in using the Green-Gauss theorem to estimate these gradients.  The crucial difference between \citet{2016MNRAS.459.1596S} and S10 is that the former estimates the gradient from the cell centers whereas S10 estimates the gradients from mesh generating points and assigns them to the cell centers.  If the mesh generating points are at the cell centers, they these methods are the same. However, this is usually not the case.  According to \citet{2016MNRAS.459.1596S}, this cell-centered gradient estimate improves this convergence properties to be comparable to the least square gradient estimate used in later versions of AREPO \citep{2016MNRAS.455.1134P}.  The gradient is then set to:
\be
\left<\grad\state\right>_i^{\rm S10} = \acharge_i^{\rm S10} \left<\grad\state\right>_i \quad\mbox{and}\quad \acharge_i^{\rm S10} = {\rm min}(1, \psicharge_{ij}),
\ee
where $\acharge_i^{\rm S10}$ is a slope limiter defined by S10 that reduces numerical oscillations near strong gradients, i.e., shocks, and $\psicharge_{ij}$ is defined as
\be
\psicharge_{ij} = \left\{\begin{array}{ll}
(\state^{\rm max}_j - \state_i)/\Delta\state_{ij} & {\rm for\ }\Delta\state_{ij} > 0 \\
(\state^{\rm min}_j - \state_i)/\Delta\state_{ij} & {\rm for\ }\Delta\state_{ij} < 0 \\
1 & {\rm otherwise},
\end{array}\right.
\ee
where $\state^{\rm max}_j$ and $\state^{\rm min}_j$ are the maximum and minimum values of the state vector among the neighbors (including itself) of the $i$th cell, and $\Delta\state_{ij} = \left<\grad\state\right>_i\cdot\left(\cmr_j - \cmr_i\right)$ is the variation computed from the gradient to the cells containing the maximum and minimum values of components of the state vector.
As noted by S10, this slope limiter is not total variation diminishing (TVD) so spurious oscillations can still occur near strong gradients.  To control this, we follow the suggestion by \citet{2011ApJS..197...15D} and (optionally) apply an additional correction when computing the gradient between cells $i$ and $j$
\be
\left<\grad\state\right>_i' = \acharge_i^{\rm DM} \left<\grad\state\right>_i^{\rm S10} \quad\mbox{and}\quad \acharge_i^{\rm DM} = {\rm min}(1, \psicharge_{ij}'),
\ee
where $\acharge_i^{\rm DM}$ is a slope limiter set between the $i$th and $j$th cells and $\psicharge_{ij}'$ is
\be
\psicharge_{ij} = \left\{\begin{array}{ll}
{\rm max}[\theta(\state_j - \state_i)/\Delta\state_{ij}), 1.] & {\rm for\ } |\Delta\state_{ij}| > 0 \\
1 & {\rm otherwise},
\end{array}\right.
\ee
where $\theta < 0.5$ gives a TVD limiter and $\Delta\state_{ij}$ is defined as above, but is now applied only to the $j$th cell and not maximized over all the neighbors. Note that $\acharge^{\rm S10}$, $\acharge^{\rm DM}$, and $\psicharge_{ij}$ are defined per component of the state vector and that setting $\theta = 1$ reduces it to the S10 slope limiter.

We note that we work with the conserved variables whereas S10 works with the primitive variables to estimate the face centered primitive variables, which is then transformed to conserved quantities.  We also note that S10 computes the face variable using only one estimate for the gradients at the n-th timestep compared to the two estimates for the gradients that we do here. We believe that the S10 method is somewhat faster, though we have believe our methodology to be somewhat more robust (against producing unphysical values) especially in regions with large gradients.


As the conserved quantities evolve or are reconstructed, the resulting density and internal energy can become unphysical, i.e., not positive definite.  This is not a new problem with Eulerian codes in general and a number of fixes must be applied.  For instance, both FLASH \citep{2000ApJS..131..273F, 2008ASPC..385..145D} and RAMSES \citep{2002A&A...385..337T} enforce a density and temperature (internal energy) floor on these quantities so that they do not become too small.  How and where these conditions are enforced produces a balance between correctness and numerical robustness.  For \changaMM, we have opted toward greater robustness (and user sanity) by enforcing a density and temperature floor in two places: at the cell center after every timestep and on the faces of the cell after the reconstruction step.

\subsection{Determining the Velocities and Timesteps of Mesh Generating Points}\label{sec:centering}

A particular feature of the scheme (as noted by S10) is its flexibility in that it operates as a Eulerian scheme, a Lagrangian scheme, or a continuous hybridization between the two.  For instance, setting the velocities of the mesh generating points to 0 ($\meshv = 0$) allows the code to operate as a Eulerian scheme or static-mesh (SM) scheme.  Likewise, setting $\meshv = \vel$ allows the code to operate as a Lagrangian scheme.  A hybrid between these two schemes would be $\meshv = \eta \vel$, where $\eta$ is arbitrary.  While this capability is available, in practice we have mainly used $\eta = 1$.

One important improvement for this scheme that was pointed out by S10 is the need to keep the Voronoi cells as regular or ``round'' as possible for which S10 proposed a method to maintain roundness.  Here we briefly describe the method in keeping with our desire that this paper serve as a complete and definitive reference for \changaMM.  As discussed in S10, ``roundness'' is maintained when the mesh generating points, $\meshr_i$, are close to the centers of mass of the Voronoi cells, $\cmr_i$. To do this, we need to correct the velocity $\meshv_i$ such that it drifts toward $\cmr_i$ at some fraction of the local sound speed $\chi c_{s,i}$. Once it is sufficiently close, this correction can be continuously reduced to zero.  The method proposed by S10 utilizes a definition of an effective radius for these cells, $R_i$ as
\be
\frac {4\pi R_i^3}{3} = V_i,
\ee
where $V_i$ is the volume of the cell.  The correction is then
\be
\Delta\meshv = \chi c_{s,i} \hat{\boldsymbol{d}}_i\left\{
\begin{array}{cl}
 0 &\mbox{for } d_i < 0.9\zeta R_i\\
  \frac{d_i - 0.9\zeta R_i}{0.2\zeta R_i} & \mbox{for } 0.9\zeta R_i \le d_i \le 1.1 \zeta R_i \\
 1 & \mbox{for } d_i > 1.1\zeta R_i
\end{array},
\right.
\ee
where $\boldsymbol{d}_i = \cmr_i - \meshr_i$ and $\zeta$ is a constant that we normally set to be $0.25$. For most of our runs, we have set $\chi = 1$ as was done in S10.

The definition of an effective radius also allows us to determine a timestep for each cell.  Currently, we have only implemented a global timestep in \changaMM, so we can define the timestep per cell as
\be
\Delta t_i = \eta_{\rm CFL}\frac {R_i} {c_{s,i} + |\vel_i - \meshv_i|},
\ee
where $|\vel_i - \meshv_i|$ is the velocity of the fluid relative to the velocity of the mesh generating point, and $\eta_{\rm CFL}$ is the Courant-Friedrich-Levy coefficient and is usually set to be $\eta_{\rm CFL} = 0.5-0.8$.  The global time step is then set to be the global minimum $\Delta t_i$ or
\be
\Delta t_{\rm cfl} = {\rm min}_i \Delta t_i,
\ee
which is then sent to the timestepper in ChaNGa.  Here, a modification that occurs is that ChaNGa is geared toward individual timesteps and determines the timestep to use to be a factor of $2^{-n}$ of a global time step $\Delta t_{\rm global}$, where $n=0$ to $30$.  Hence the timestep that is used is
\be
\Delta t = 2^{-n} \Delta t_{\rm global} \quad\mbox{where } 2^{-n} \le \frac{\Delta t_{\rm cfl}}{\Delta t_{\rm global}} \le 2^{-(n-1)}.
\ee
The individual timestepping scheme implemented in ChaNGa allows for the possibility of extending \changaMM\ to individual timesteps which should greatly speed up problems with large dynamic ranges.  We will discuss this future improvement in \S~\ref{sec:future}.

\subsection{Entropy versus Energy Evolution}\label{sec:entropy}

The numerical solution to the energy equation (\ref{eq:energy}) can be subject to spurious numerical heating particularly in cold flows that are dominated by a (numerically) noisy gravitational field.  To help remedy this issue, S10 introduced an entropy equation that is absent a source to alleviate this spurious heating:
\begin{equation}
 \ddt{\rho s} + \grad\cdot{\rho s \vel} = 0\label{eq:entropy},
\end{equation}
where $s$ is the specific entropy. The solution to equation (\ref{eq:entropy}) follows the same strategy as in equation (\ref{eq:state}) with the addition a column of $\rho s$ to the state vector, $\state$, $\rho s \vel$ to the flux function, $\flux$, and $0$ to the source function, $\source$. The energy evolution equation (\ref{eq:energy}) and entropy evolution equation (\ref{eq:entropy}) are redundant.  Moreover, they are inconsistent with each other; the energy equation (\ref{eq:energy}) has a source whereas the entropy equation (\ref{eq:entropy}) does not.  This redundancy and inconsistency is not a bug but rather a feature.  In particular, the entropy equation (\ref{eq:entropy}) is not subject to spurious heating whereas the energy equation (\ref{eq:energy}) is.  On the other hand, the energy equation can capture the effects of source heating and shock heating whereas the entropy equation cannot.

To allow for the capture of external and shock heating, while minimizing spurious heating, S10 switches the evolution of the gas between the energy and entropy equations in a cell depending on the existence of shocks of sufficient strength.  In particular, S10 determines the maximum Mach number of the shocks (if any) over all the faces of the cell (during the solution of the Riemann problem).  If this maximum mach number exceeds a certain threshold, $\mathcal{M}_{\rm th} = 1.1$, then the energy equation is used.  Otherwise the entropy equation is used.  The internal energy or entropy is then reset based of the particular equation that was integrated.  Hence, the noninclusion of a source term in the entropy equation is intentional, i.e., we eliminate spurious heating due to numerical noise in the gravity solve.  S10 demonstrated the utility of this switch in the Santa Barbara cluster test (their Figures 44 and 45), where the use of this switch prevent artificial high redshift heating of the gas, albeit with a scheme that effectively used a much larger $\mathcal{M}_{\rm th}$.

We have implemented this same scheme in \changaMM, and we have chosen the same values as in S10 with $\mathcal{M}_{\rm th} = 1.1$.  We note that we can choose to use only the energy evolution equation (\ref{eq:energy}) if we set $\mathcal{M} = 0$.  We also note that other moving mesh codes including TESS\citep{2011ApJS..197...15D}, FVMHD3D\citep{2012ApJ...758..103G}, ShadowFax\citep{2016A&C....16..109V}, RICH\citep{2015ApJS..216...35Y}, and DISCO\citep{2016ApJS..226....2D} do not include the entropy equation.

\subsection{Initialization}\label{sec:initialization}

In our implementation of \changaMM, we have kept the input/output and timestepping architecture of ChaNGa so that we can analyze the resulting output using the large library of already developed analysis software such as YT \citep{2011ApJS..192....9T}, which is our analysis and visualization platform of choice.  It also allows us to rapidly switch between a MM (or SM) solver and an SPH solver with the inclusion of a single configuration flag and use the exact same initialization files to facilitate easy comparison between the different solvers.

That being said, there are some significant differences between the amount of information being needed by the SPH solver and needed by the MM solver.  In SPH, the conserved quantities per particle are stored including mass, momentum (or velocity), and internal energy.  Primitive quantities like density and pressure are derived by using a kernel to smooth over a few neighbors.  In the MM formalism, these primitive quantities are first-class quantities that are stored per cell.  Hence, when we store outputs from \changaMM\ we also store these primitive quantities as well.  The structure of the ChaNGa/Gasoline output data format allows the analysis software to deal with these quantities appropriately.

As a result, \changaMM\ has two initialization modes.  It can be initialized using SPH initialization files used for ChaNGa/Gasoline or it can be initialized using \changaMM\ initialization files.  In the former, we use the particle positions from the SPH initialization files as our mesh generating points, and we set the density using the SPH smoothing kernel on the particle data.  For the other primitive variables, velocity and internal energy, we use the quantities in the SPH particle.  We then update the conserved quantities of each cell and proceed with the solve.  In the latter, when \changaMM\ initialization (or previous output) files are available, we use the primitive quantities that are stored directly to initialize the mesh-generating points.  As a result, one can restart from an output using the exact conditions as when it was outputted.

As with all Eulerian or ALE schemes, the mesh generating points must span all of space and the space has to be well defined, i.e., compact with appropriate boundary conditions.  This is unlike the case in SPH where space can be effectively infinite.  As a result the initialization files must contain additional points that span all of space, which is of fixed size with defined boundary conditions.  Currently, we have only implemented periodic boundary conditions, though we may implement reflecting boundary conditions in the future.   Here, we must be explicit with these additional points, which is in contrast to AREPO, where these points are generated automatically by filling in the empty leaf cells of a Barnes-Hut tree.  This precludes us from using the exact same initial conditions that have previously been developed for ChaNGa/Gasoline, but a translator will be developed in the near future as described in \S~\ref{sec:future}. For now, these additional mesh generating points have to be specified explicitly.

\section{Test Problems}\label{sec:test problems}

We validate \changaMM\ with a few test problems and compare the results against those obtained from the SPH module using identical initial conditions.  The Voro++ library is restricted to 3-D, so all the tests here are carried out in 3-D.  In the case where the problems are fundamentally 1-D or 2-D, we use boxes with appropriate aspect ratios to emulate this effect.  In principle, it is possible to (trivially) modified the construction of the Voronoi tessellation to exactly produced 1-D or 2-D problems, but we have not chosen to do so.  These test problems used in our validation are the Sod shock-tube problem, the Sedov-Taylor point explosion, the Gresho-Chan vortex problem, the Evrard collapse problem, a star in hydrostatic balance, and a stellar merger problem.

For these test problems, we use SPH initialization conditions, which we generate as follows. We begin with a glass of 4,096 particles in a unit cube and generate a $n^3$, where $n=16$, block of unit cubes each containing the glass.  We can then give each particle a mass and rescale the size of the block to be an appropriate representation of these initial conditions.  After rescaling, we cut down the block to produce cube of the appropriate size that can be fed to \changaMM\ to solve either in SPH or MM mode.  We note that our current initialization limits us to particle numbers of $\approx 16$M particles, though this can be trivially increased by using a larger $n$. We note that while the Sedov-Taylor and Evrard problem are scale-free, we have chosen to pick a specific (but arbitrary in cgs) scale for the initial conditions.  For the Sod shock-tube and Gresho-Chan vortex problems, these are in normalized units.

For the SPH solver in the test problems below, we have set the number of neighbors to be 64 and have used the standard $M_4$ kernel\citep{1992ARA&A..30..543M,2005RPPh...68.1703M}.  Additionally, we have used the Balsara switch to suppress the shear viscosity with default $\alpha$ and $\beta$ values ($\alpha = 1$ and $\beta=2$) \citep{1995JCoPh.121..357B}.

\subsection{Sod Shock-tube Test}

We begin with a the simple Sod shock-tube which is initialized with a periodic box (in normalized units) of dimension (16, 1, 1) centered at (0,0,0) with 41K mesh generating points.  We initialize the $\gamma=1.4$ gas on left side of the box ($x<0$) with $\rho = 1$ and $P=1$ and on the right side ($x>0$) with $\rho=0.25$ and  $P = 0.1795$, which are the same parameters used for a number of other code test including Gasoline \citep{2004NewA....9..137W} and AREPO (S10). The gas is initially stationary ($v=0$).

\begin{figure*}
 \includegraphics[width=0.49\textwidth]{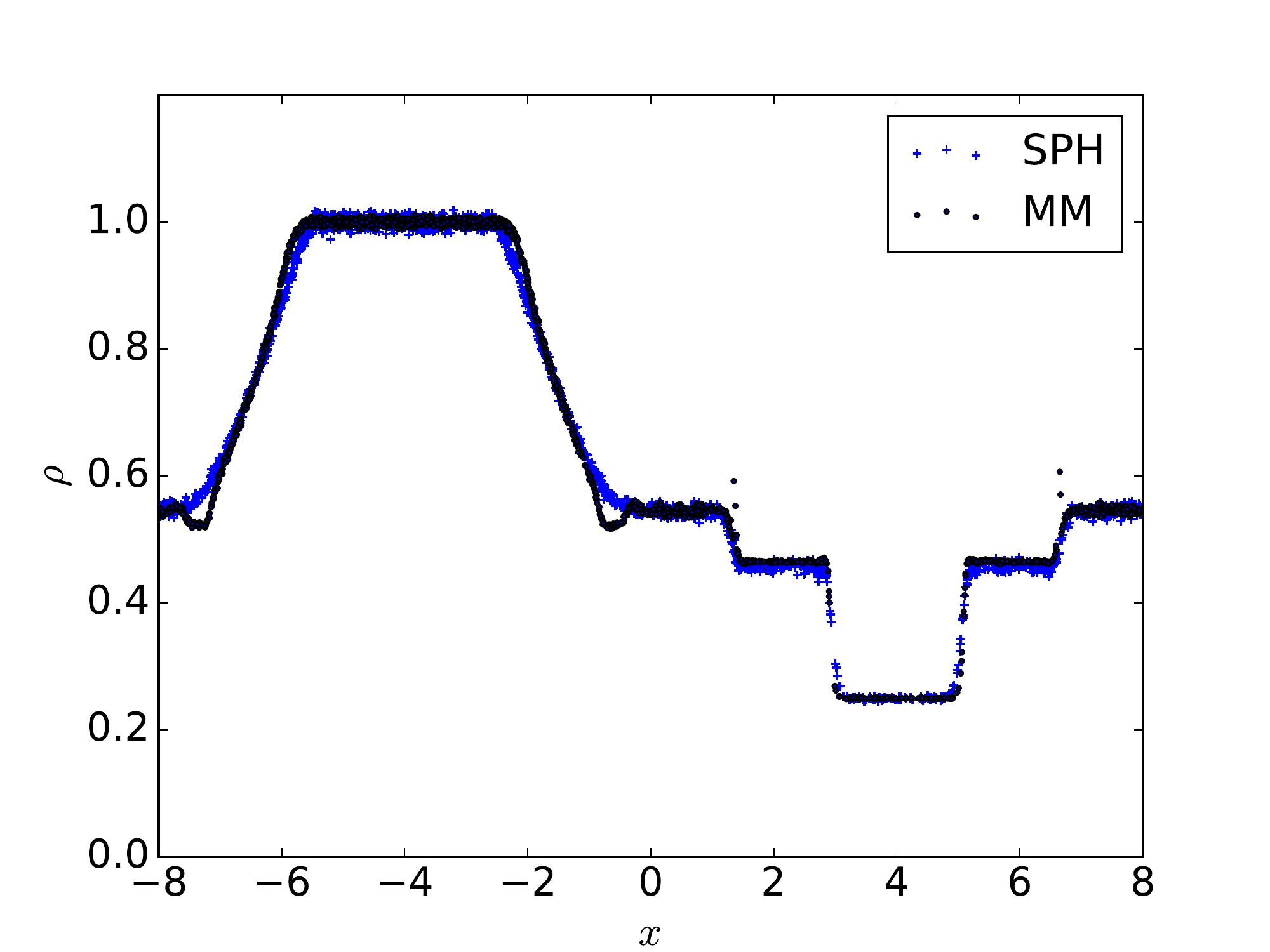} \includegraphics[width=0.49\textwidth]{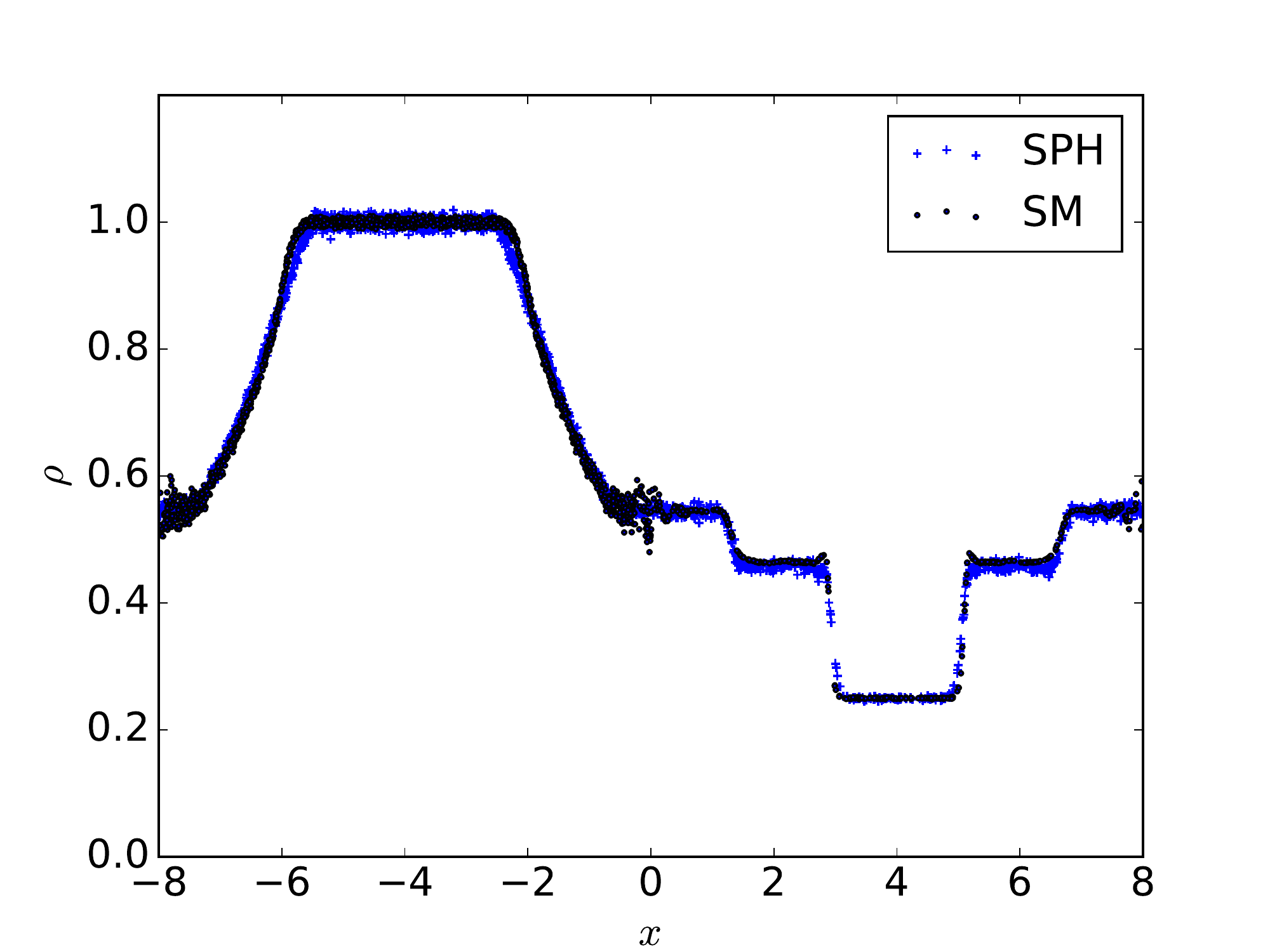}\\
  \includegraphics[width=0.49\textwidth]{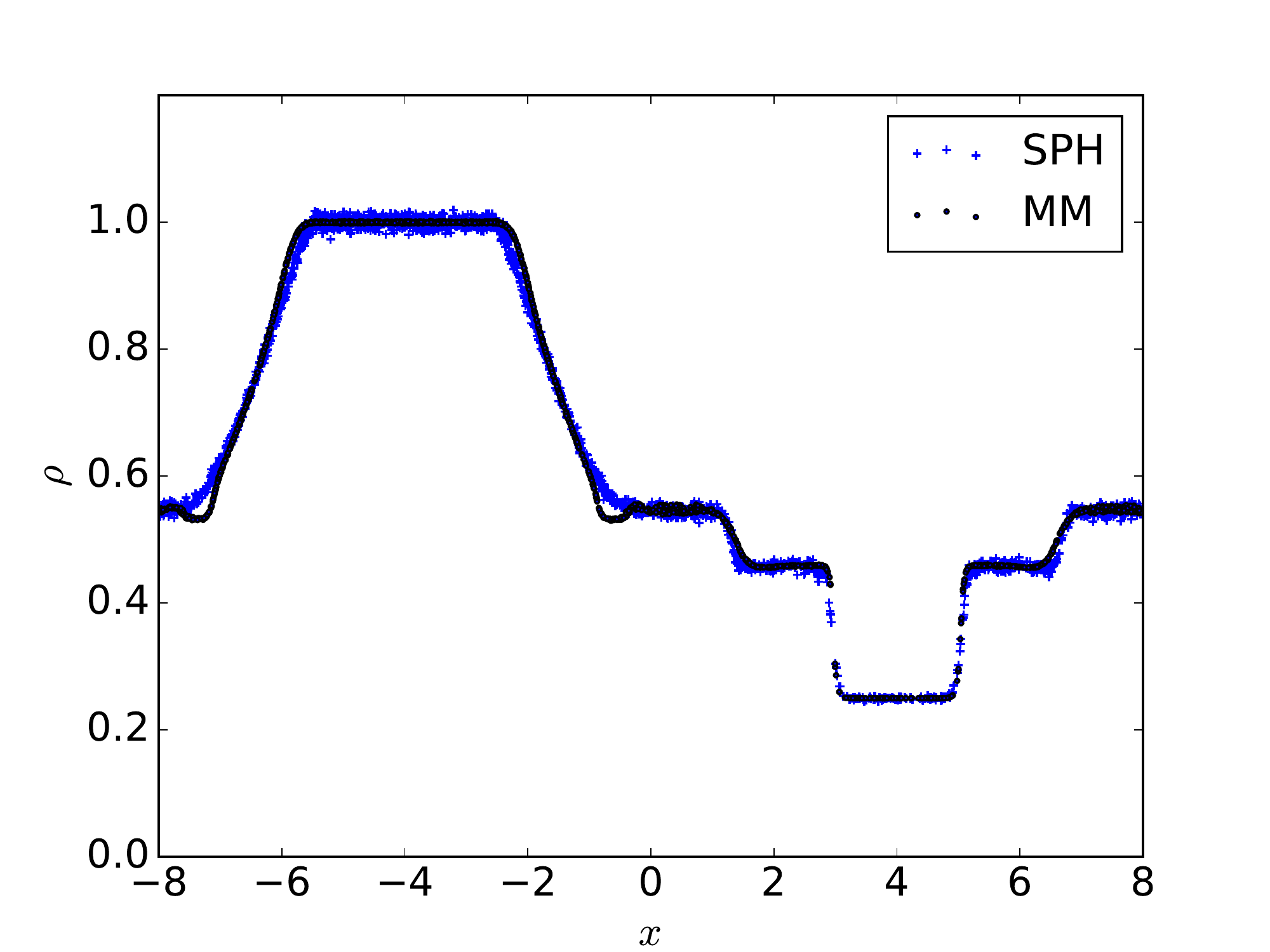} \includegraphics[width=0.49\textwidth]{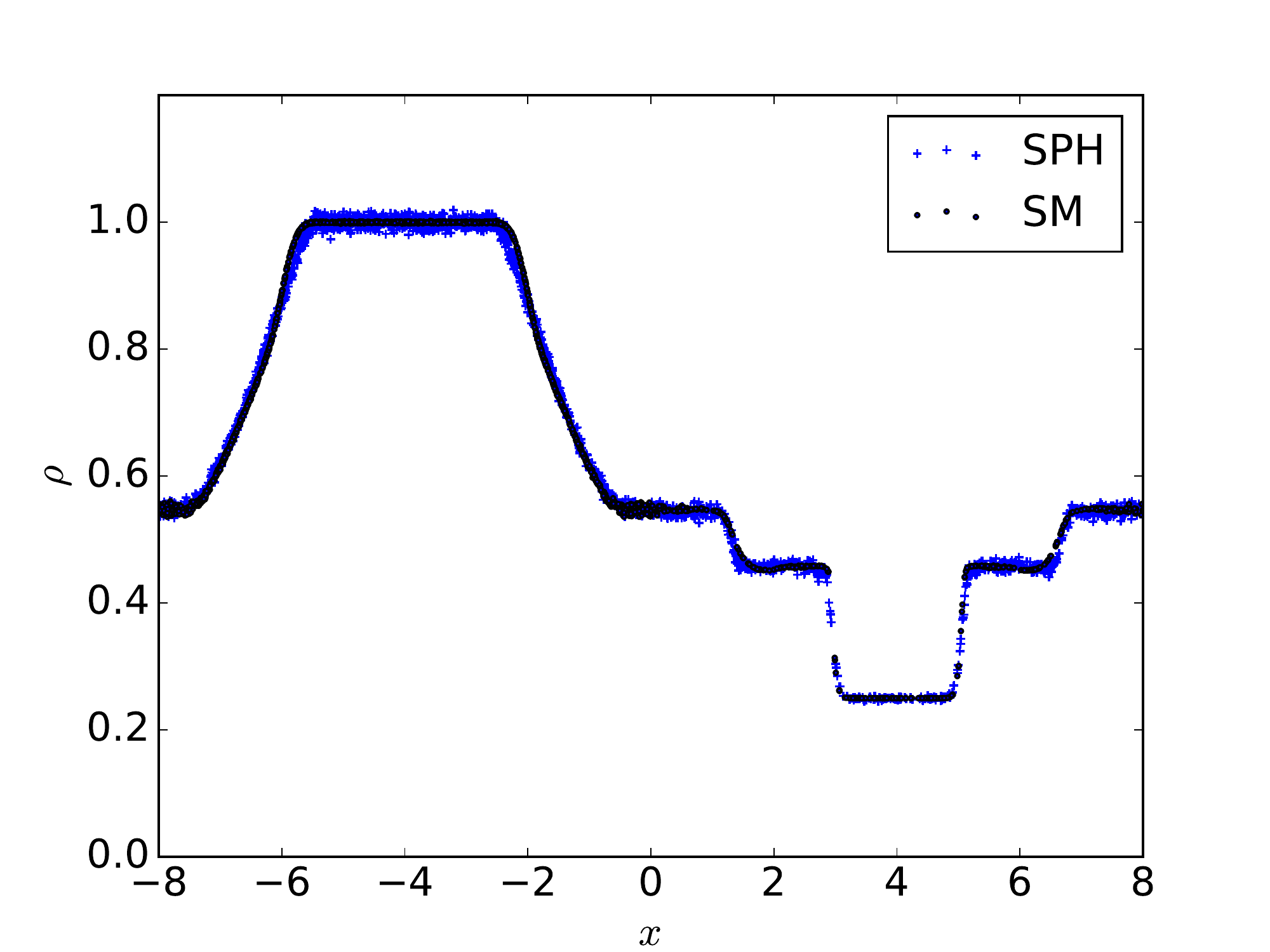}
   \caption{ Density as a function of position at $t=2$ for the Sod shocktube test solved using SPH (blue crosses), MM (left plot, black dots), and SM (right plots, black dots) using the HLLC solver (top plots) and the HLL solver (bottom plots).\label{fig:sod-profile}}
\end{figure*}

In the top plots of Figure \ref{fig:sod-profile}, we plot the density (downsampled by 10) at $t=2$ for the three solvers that we have in MANGA: the SPH solver, the MM solver, and the SM solver using the HLLC solver and $\theta=0.49$. The slope limiter is TVD, but barely so, and thus the spurious oscillations are evident in both the SM and MM cases.  This can be remedied in part by decreasing $\theta$ so that it become more diffusive, but we have found that using the HLL solver, which is more diffusive, eliminates these oscillations. This is shown in the bottom plots of Figure \ref{fig:sod-profile} where these oscillations are eliminated in the SM case and reduced significantly in the MM case.

The MM and SM solvers are roughly comparable to the SPH solver.  At this high particle count, the discontinuities are equally well resolved by SPH, MM, and SM solvers.  However, there is less variation in the per cell information in the MM and SM case compared to SPH.  This is seen in the spread of density for the SPH solver in the regions where the density should be constant.  This is due to the nature of the density estimator in SPH, which is subject to the locally random distribution of particles.

\subsection{Sedov-Taylor Explosion}

\begin{figure*}
 \includegraphics[width=\textwidth]{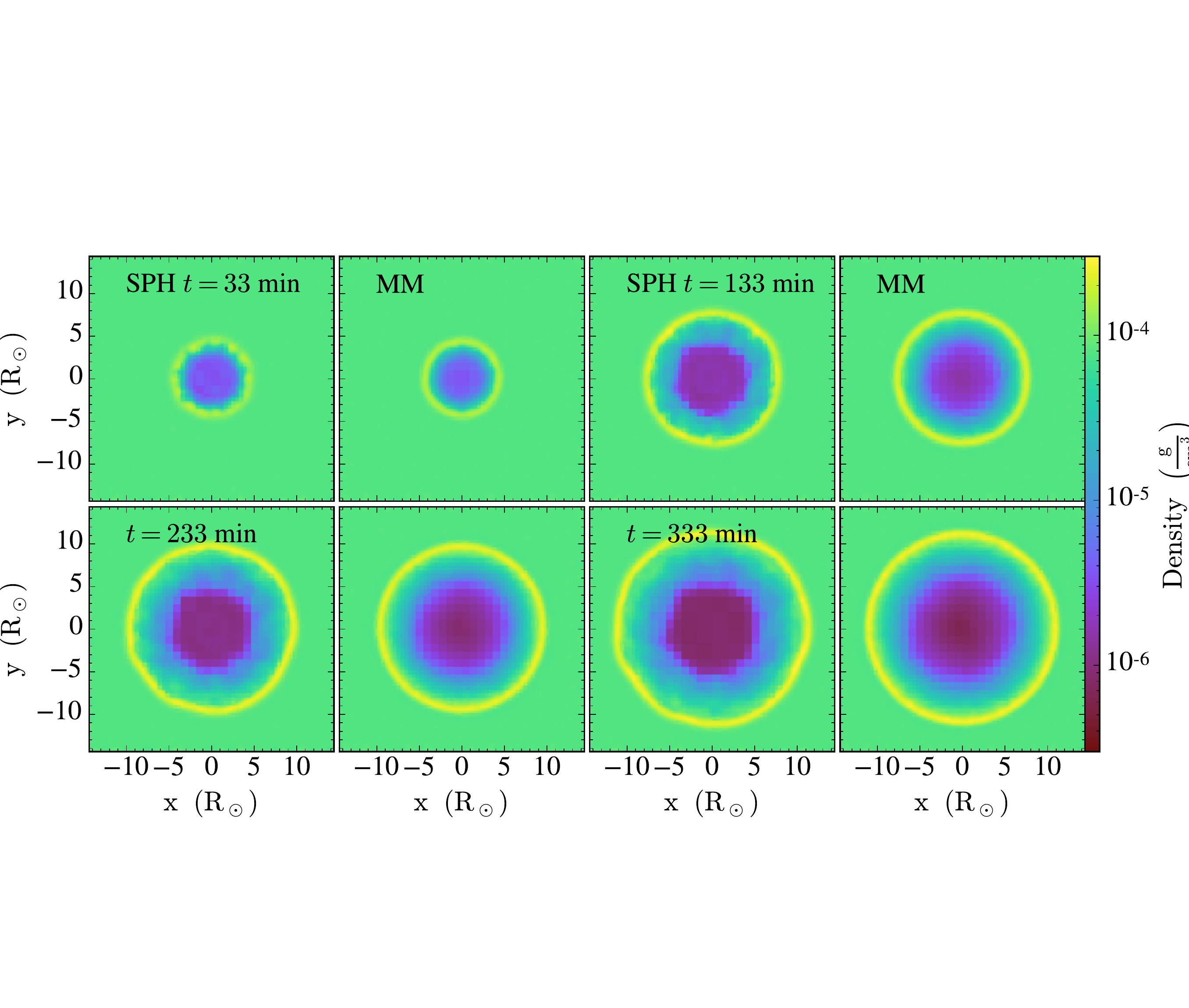}
   \caption{Slice plots of a point explosion that consists of a $1\,M_{\odot}$ box of pure hydrogen gas in a $2\times 10^{12}$ cm periodic box.  We show the evolution of the explosion at times,  $t=33$, $133$, $233$, and $333$ minutes using the MM and SPH solvers. \label{fig:sedov}}
\end{figure*}

In this test problem, we initialize a uniform periodic box with linear size $2\times 10^{12}$ cm that contains $1\,M_{\odot}$ of pure hydrogen gas at an initially constant density, $\rho = 2.5\times 10^{-4}\,{\rm g\,cm}^{-3}$.  In the central $r < 8.7\times 10^{10}$ cm, we initialize the temperature to be $T=10^9$ K, while the remainder of the gas is initialized to be $10^4$ K. We use a total of 255K mesh-generating points in this example.  This results in a point explosion that is described by the Sedov-Taylor solution. In Figure \ref{fig:sedov}, we show the evolution of this explosion at times, $t=33$, $133$, $233$, and $333$ minutes in the MM and SPH mode.

\begin{figure}
 \includegraphics[width=0.5\textwidth]{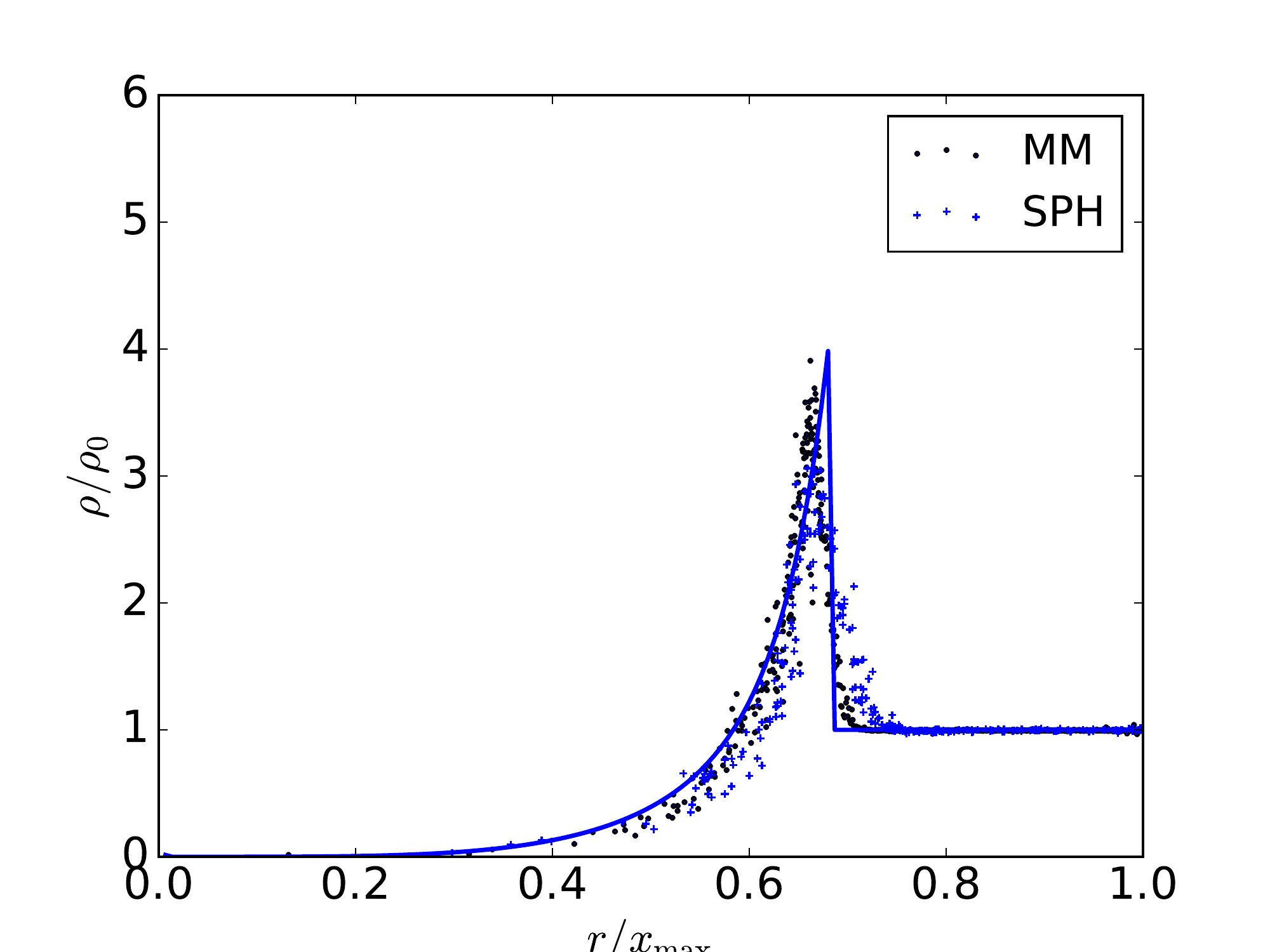}
   \caption{ Profile plots of a point explosion shown in Figure \ref{fig:sedov} at $t=233$ minutes for both the SPH and MM solvers. The analytic solution is shown by the solid blue line. \label{fig:sedov-profile}}
\end{figure}

In Figure \ref{fig:sedov-profile}, we show the radial profile of the density as a function of radius at the same times.
For comparison, we also compute the same test problem with the same initial conditions using SPH and fixed grid mode and plot these radial profiles.  The resolution of the shock is slightly better in MM mode compared to SPH in that the peak density is reached. But at the resolution in this simulation, the superior resolution of Godonov methods to resolve shocks compared to SPH is not evident. Visually, the resolution and evenness of the shock looks better in Figure \ref{fig:sedov}, where particle effects in the density estimator in YT is evident in the SPH case, but nearly absent in the MM case. Essentially, this is due the mass of a mesh generating point being tied to the volume that the Voronoi cell which it defines as oppose to the fixed mass of an SPH particle.

\subsection{Gresho-Chan Vortex Problem}

The Gresho-Chan vortex \citep{Gresho1990} is an inviscid, cylindrical vortex. The initial state is steady, with centrifugal acceleration balancing the pressure gradient.  The simulation volume is periodic with unit dimensions in $x$ and $y$ and thinner in the $z$ direction.  While the test is technically 2D, because we use glass initial conditions the ability to retain translation symmetry in $z$ is a test of the method.  The density is uniform $\rho=1$ and the pressure, $P$, as a function of radius, $r$, is given by,
\begin{equation}
    P(r) =
    \begin{cases}
        5+12.5\,r^2 & \quad (0 \leq r < 0.2)\\
        9+12.5\,r^2-20\,r+4\ln{5r} & \quad (0.2 \leq r < 0.4)\\
        3+4\ln{2} & \quad (r \geq 0.4)\\
    \end{cases}
\end{equation}
with a matching tangential velocity function,
\begin{equation}
    v_{\phi}(r) =
    \begin{cases}
        5r & \quad (0 \leq r < 0.2)\\
        2-5r & \quad (0.2 \leq r < 0.4)\\
        0 & \quad (r \geq 0.4).\\
    \end{cases}\label{eq:gresho_v}
\end{equation}
and $v_r=0$, $v_z=0$.
We evolve the vortex to $t=2$, or $\sim 1.6$ rotations of the
peak.  We run a simulation with $\sim 50$ K particles in a flattened periodic box with unit length in x and y and a length of 0.2 in z.  While the box is 3D, it gives an equivalent 2D resolution of $64^2$.

\begin{figure}

\includegraphics[width=0.49\textwidth]{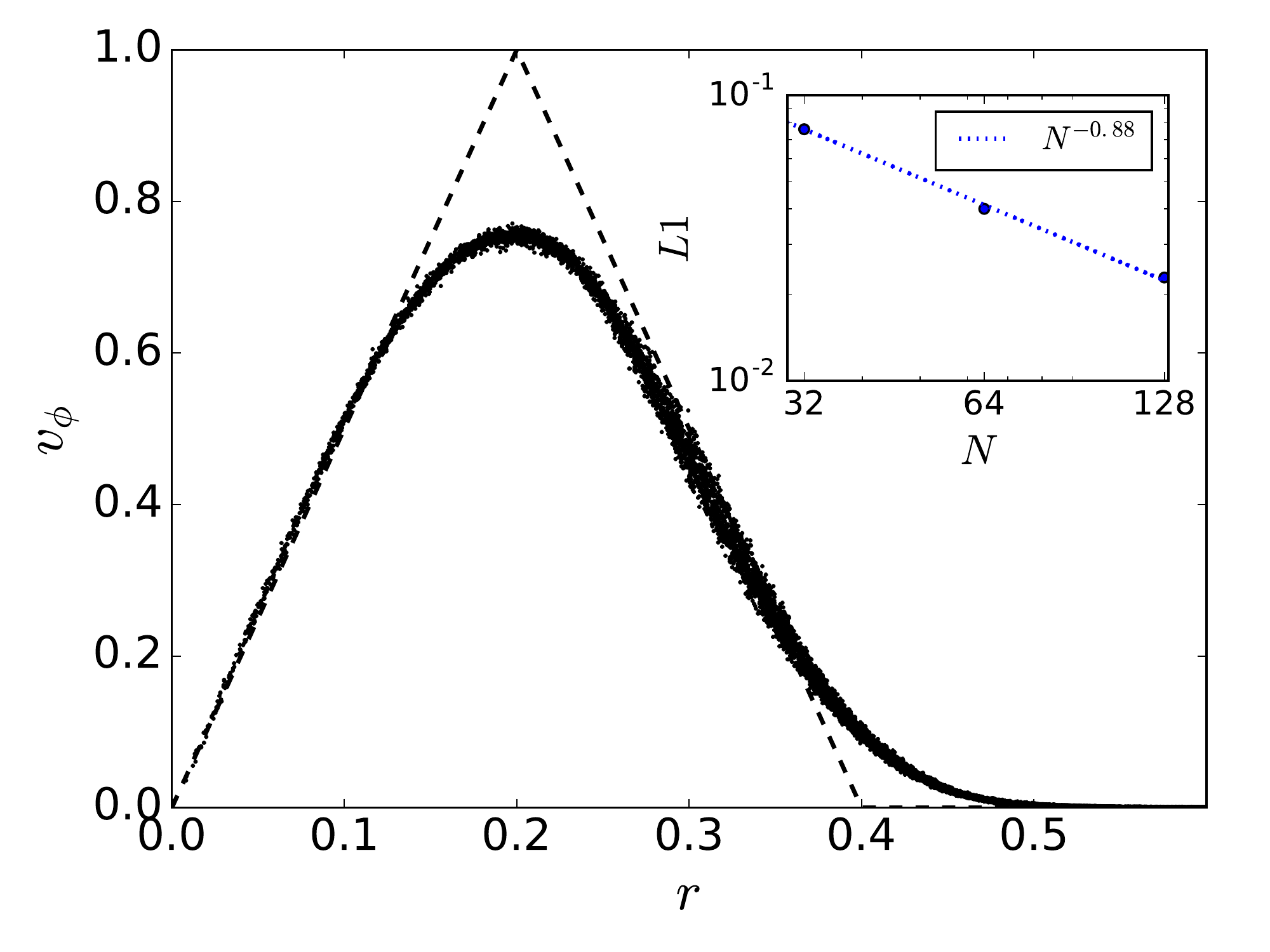}\\

   \caption{ Tangential velocity, $v_{\phi}$ as a function of radius, $r$, for a Gresho vortex for a equivalent 2-d resolution of $64^2$ at $t=2$, which corresponds to 1.6 vortex rotation at the most rapid spinning point at $r=0.2$.  For reasons of clarity the points have been downsampled by a factor of 10.  The analytic (stable) solution is shown by the dashed line. Inset plot: L1 error norm as a function of effective 2D resolution, $N^2$, for $N=32$, $64$, and $128$.  Also shown is the scaling $L1\propto N^{-0.88}$ by the dotted line.\label{fig:gresho-profile}}
\end{figure}

In \corr{the main plot of} Figure \ref{fig:gresho-profile}, we show the tangential velocity, $v_{\phi}$, as a function of radius, $r$, from the center of the vortex at $t=2$, which corresponds to $\sim 1.6$ vortex rotation at $r=0.2$.  We do not make a comparison with SPH as the  SPH solver in ChaNGa v 3.2 suffers from large angular momentum diffusion and is unable to solve this problem with any fidelity.  More modern implementation especially those with artificial viscosity limiters are much better at integrating the Gresho vortex \citep{Wadsley2017}.
In any case, \changaMM\ does a reasonable job of replicating the analytical solution (eq.(\ref{eq:gresho_v})) shown as the dashed line.  \corr{In the inset plot of Figure \ref{fig:gresho-profile}, we plot the L1 error norm, $L1$, of the tangential velocity, $v_{\phi}$, as a function of the equivalent 2D resolution, $N^2$, for $N=32$, $64$, and $128$.  The L1 error norm, $L1$, is given by:
\begin{equation}
 L1 = \frac 1 N_{\rm pts} \sum_i |v_{\phi,i} - v_{\phi}(r_i)|, 
\end{equation}
where $N_{\rm pts}$ is the number of mesh-generating points, $v_{\phi,i}$ and $r_i$ are the fluid tangential velocity and distance from the origin associated with that point, and 
$v_{\phi}(r)$ is the analytic solution given by equation (\ref{eq:gresho_v}). The dotted line in the inset plot shows scaling of the L1 error norm with $L1 \propto N^{-0.88}$, which is similar to the scaling L1 error norm of the Gresho vortex ($\propto N^{-0.8}$) found using the modern SPH code Gasoline2 \citep{Wadsley2017}.  Convergence is worst than the L1 scaling found in S10, but S10 ran a 2D test whereas our test is in 3D.     } 

\subsection{Evrard Collapse Problem}

In this test problem, we initialize a $m_0 = 1000\,M_{\odot}$ sphere with a mass profile given by
\begin{equation}
 m(r) = {m_0}\left(\frac{r}{r_0}\right)^{2},
\end{equation}
where $r_0 = 1.7\times 10^{12}$ cm. This gives a $\rho \propto 1/r$ profile.  We embed this sphere in periodic box with linear size $2\times 10^{12}$ cm and remove the excess that extends beyond this domain.  We use a total of 450K mesh-generating points in this simulation. The gas temperature is set to a value of $T=1000$ K, which is small compared to the gravitational binding energy.  As this gas evolves, it will initially fall toward the center, produce a shock that runs outward before finally settling to a hydrostatic equilibrium.

\begin{figure*}
M \includegraphics[width=0.8\textwidth]{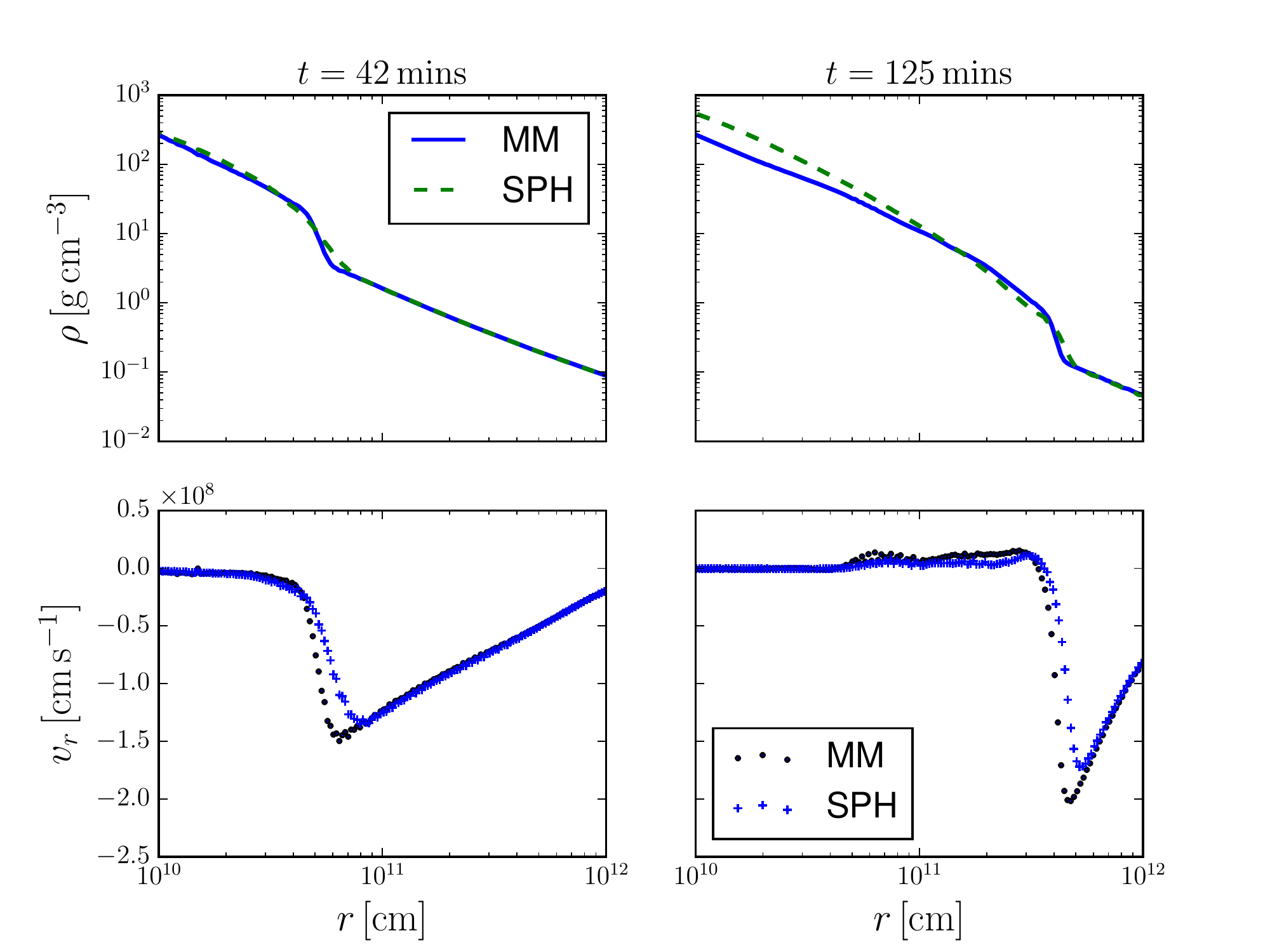}
   \caption{Snapshots of the density and radial velocity profile for the Evrard collapse test at $t=42$ and $125$ minutes for both the SPH and MM solvers.  The two methods agree on the position of the shock though the MM code appears to better resolve the shock.\label{fig:evrard-profile}}
\end{figure*}

We show the evolution of the density and radial velocity profile in Figure \ref{fig:evrard-profile} at two times: $t= 42$ and $125$ mins. The SPH and MM solvers produce essentially the same shock position though it does appear that the shock in the MM is better resolved.  This also appears to be the case in the low resolution studies in S10.

\subsection{Stellar Hydrostatic Balance}

As another test of gravity, we investigate the ability of \changaMM\ to model a hydrostatic star for many dynamical times.  We create a star that is modeled as a purely ionized hydrogen $n=3$ polytrope that has a central temperature of $T_c=10^7$ K and total mass $M=1\,M_{\odot}$.  We solve hydrostatic balance in 1-D:
\begin{equation}
  \frac{\partial P}{\partial r} = -\frac{G M(<r)\rho}{r^2},
\end{equation}
where $M(<r) = \int_0^r 4\pi r'^2\rho(r') dr'$ is the mass contained within $r$.  Here we adopt a polytrope equation of state, $P=P_c(\rho/\rho_c)^{(1+n)/n}$, where $\rho_c$ and $P_c = \rho_ck_BT_c/m_p$ are the central density and pressure respectively, and $n$ is the index.
The resulting star has a radius of $R\approx 2\times 10^{11}$ cm.  We then use these 1-D models to create initial conditions for \changaMM\ using $10^5$ mesh generating points to model these stars.  Surrounding the polytrope is a low density atmosphere with $\rho = 10^{-6}\,{\rm g\,cm}^{-3}$, which is low compared to the central density of the polytrope ($11\,{\rm g\,cm}^{-3}$).  The total number of mesh-generating points in this simulation (star and atmosphere) is 160K.

\begin{figure*}
 \includegraphics[width=0.8\textwidth]{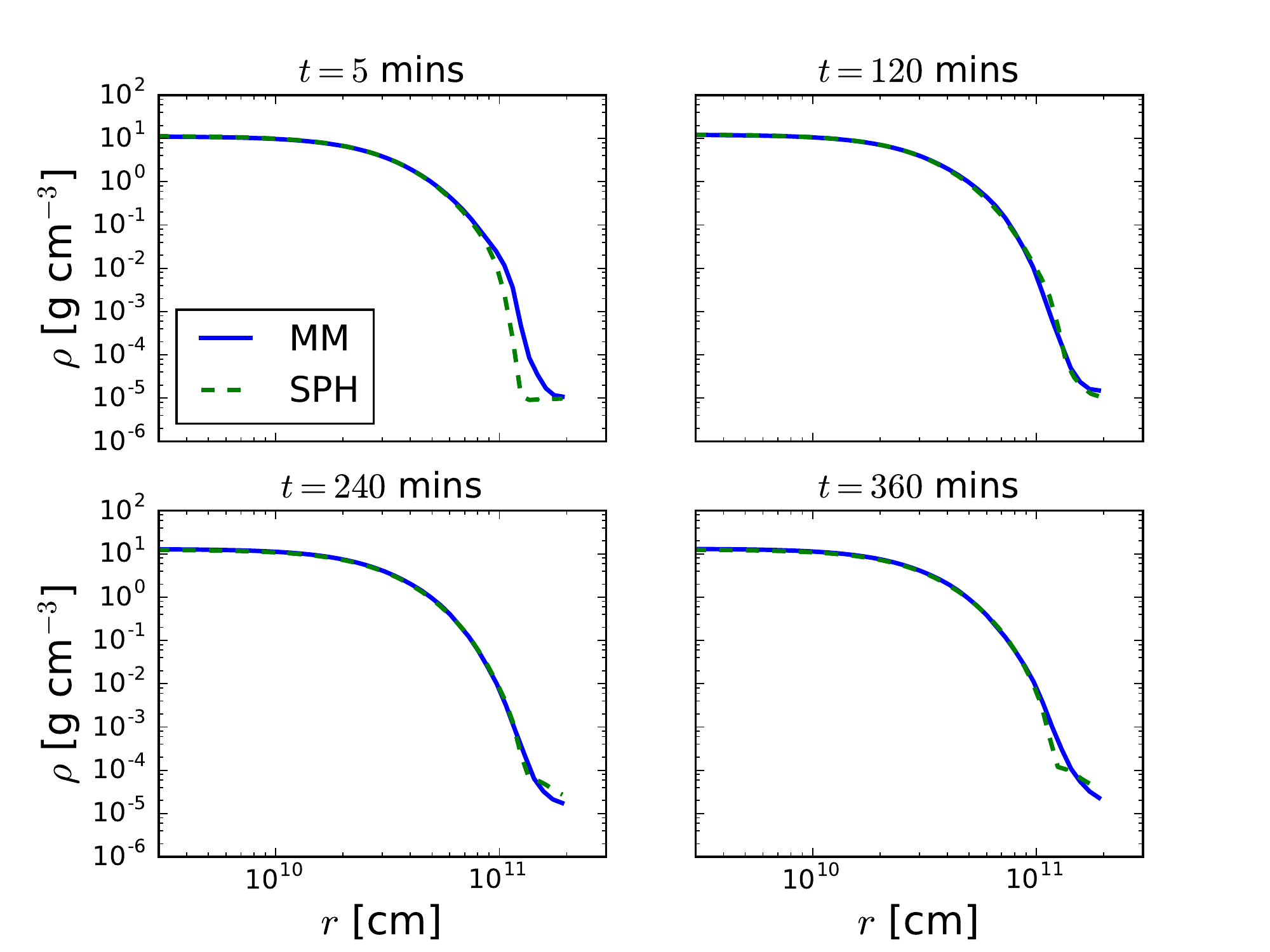}
   \caption{Comparison of the 1-D profile of a radiative solar-type star ($n=3$ polytrope) in hydrostatic balance between SPH and \changaMM\ for times in two hour intervals (0, 2, 4, 6 hours).  Note that the bulk of the star agrees between the two methods with small deviations at low densities where the strong density gradients are poorly resolved (especially in SPH).    \label{fig:star}}
\end{figure*}

We evolved the initial conditions using the SPH and MM modules in \changaMM.  It is well known that these hydrostatic models in 1-D are not perfectly hydrostatically balanced when mapped to a 3-D situation.  Rather, these stars tend to oscillate, and only with velocity and energy damping do they reach a new hydrostatic equilibrium.  We do not include the damping in this case and, instead, allow the star to oscillate freely. In Figure \ref{fig:star}, we plot the evolved star (in SPH and MM) at $t = 5, 120, 240,$ and $360$ minutes after initialization.  Here we see that the behavior of the two stars are quantitatively the same, especially for the bulk of the mass.  Only in the outer envelope is there is a departure between these two solutions, with the MM algorithm being smoother.  In any case, this demonstrates that the MM algorithm can successfully maintain hydrostatic equilibrium and that this equilibrium is similar to that of produced from an SPH code. The ability of \changaMM\ to produce hydrostatic models that agree with its SPH counterpart suggests that the MM algorithm can be applied to dynamical stellar problems.

\subsection{Stellar Merger}

\begin{figure*}
 \includegraphics[width=\textwidth]{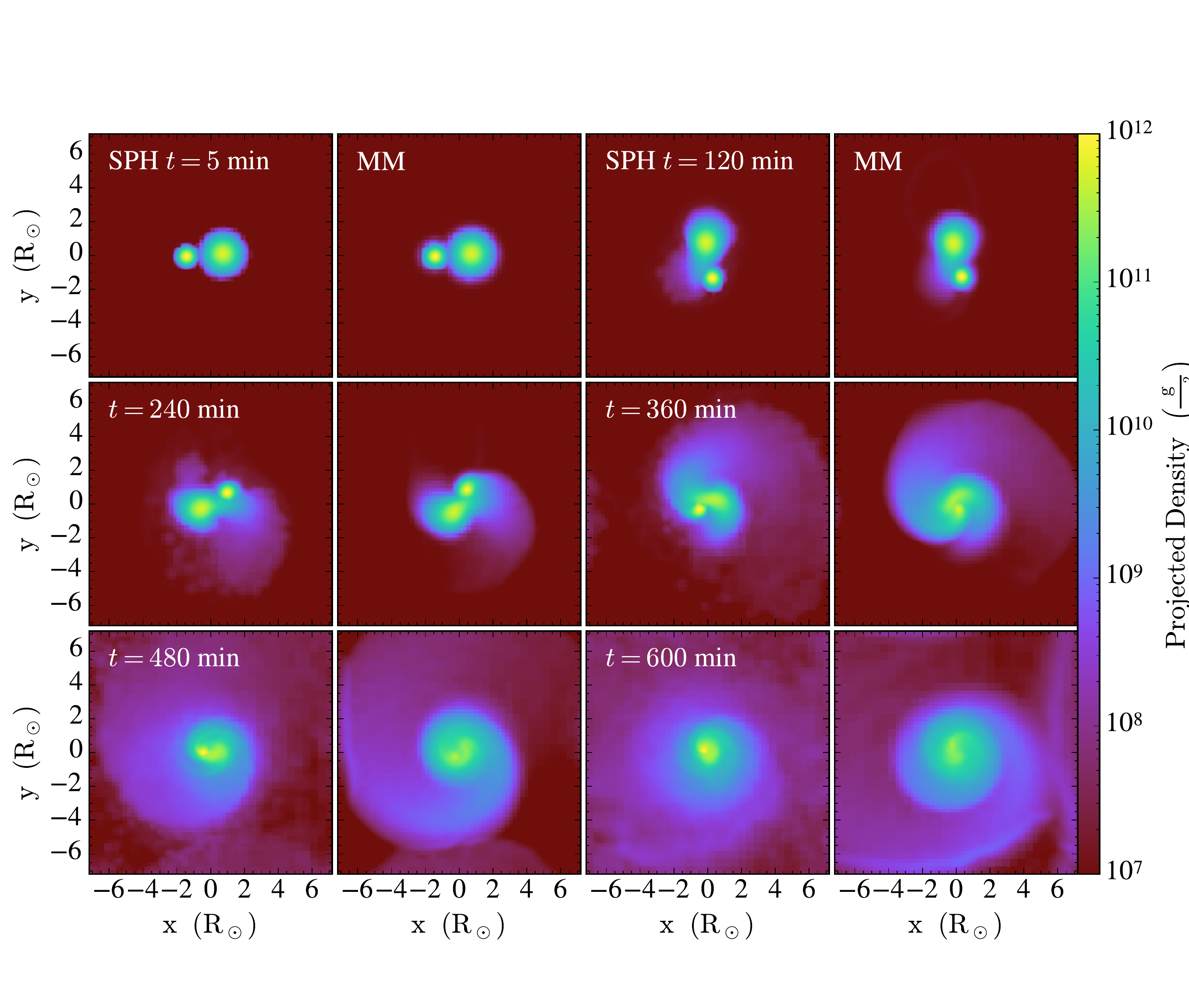}
   \caption{Snapshots of a merger between a 0.5 and 1 solar mass $n=3$ polytropes (stars) using the SPH and MM solvers at 2 hour intervals up to 10 hours.  Both solvers qualitatively produce similar results with a few notable differences.  The SPH solver produces ejecta that is clumpy compared to the ejecta produced by the MM solver.  This is likely due to numerical surface tension that is prevalent in SPH without the pressure fix.  Additionally, the smaller, denser companion is less well mixed and retains a higher density core in the SPH solver compared to the MM solver, which is also likely due to the same effect.  \label{fig:merger}}
\end{figure*}

As a test of \changaMM\ to investigate dynamical stellar problems, we consider the merger of a $m_2 = 0.5$ and $m_1 = 1 M_{\odot}$ stars, i.e., $n=3$ purely ionized hydrogen polytropes, that are initially in contact with each other, i.e., we set their initial separation at $a=r_1 + r_2$, where $r_1$ and $r_2$ are the size of star 1 and 2 respectively.  As for the stellar hydrostatic test, we include a low density atmosphere with $\rho = 10^{-6}\,{\rm g\,cm}^{-3}$.  The total number of mesh-generating points between the two stars and atmosphere is 285K. We also give them a relative velocity of $v=\sqrt{G(m_1+m_2)/a}$ so that they would be circular orbits if they were point masses.  We show their evolution at 2 hour intervals in in Figure \ref{fig:merger} up to 10 hours.

While the SPH and MM calculations look broadly the same, they do differ in the details.  For instance, the SPH models show much greater clumpiness in the disk that surrounds the star and a less well-defined (though still obvious) spiral pattern. On the other hand, the MM models are much smoother and show a long lasting spiral pattern.   The SPH models also show less disruption of the dense core of the companion, whereas it appears that the MM models appear more well mixed.  Both of these effects may be due to a spurious surface tension effects at large density gradients that are known to exist in the density formulation of SPH \citep{2007MNRAS.380..963A}.  These effects can be remedied within standard SPH by adopting different formulations based on the pressure or internal energy \citep{2013ApJ...768...44S,2013MNRAS.428.2840H} or the geometric density average used in more recent versions of ChaNGa or Gasoline\corr{2 \sout{(Wadsley Keller and Quinn,2017, submitted)}\citep{Wadsley2017}}, or alternately using a MM algorithm. 

\corr{On the other hand, \changaMM\ is inferior to the SPH solver in conserving angular momentum.  For instance, from $t=5$ to $t=240$ minutes, \changaMM\ suffers a 7\% change in the angular momentum while the SPH solver suffers only a 0.1\% change in the angular momentum.  This is due to the use of the cell's geometric quantities following S10 at the beginning of the time-step, i.e., face areas, which breaks second-order time integration. As discussed below, \citet{2016MNRAS.455.1134P} explicitly identified this problem as responsible for the non-conservation of angular momentum in white dwarf merger problems and pointed out means to remedy it.  One of the future improvements to \changaMM\ will be to remedy this problem as we discuss below.}  

\section{Discussion and Future Directions}\label{sec:future}

\changaMM\ as currently implemented can be used to study some problems of scientific interest, especially stellar mergers.  However, ChaNGa and Gasoline have mainly been used to study galaxy formation and cosmological structure formation and thus have additional physics implemented that is currently missing from \changaMM.  In addition, ChaNGa and Gasoline have implemented a few algorithmic improvements in SPH including individual timesteps and asynchronous gravity and SPH calculations to improve performance.  We plan to target a number of improvements in the near, medium, and long term to include a few of these improvements and physics to allow \changaMM\ to be more widely applied.  The improvements in order of priority (determined by some combination of utility and simplicity) are:
\begin{enumerate}
  \item \corr{{\bf True Second Order Time-integration:} \changaMM\ as currently implemented is nearly second order in time except for accounting for the changes in the mesh geometry at the half-time-step.  Instead, we follow the original implementation detailed in S10 where we use the mesh geometry at the beginning of a timestep to compute the area of the faces between adjacent cells, not accounting for the changes in the area as these cells move over a timestep.
This was identified in \citet{2016MNRAS.455.1134P} as the root cause for the violation of the angular momentum conservation in white dwarf merger problems.  There are two ways to fix this.  One method is the follow the algorithm implemented in TESS \citep{2011ApJS..197...15D} and do a (re-)construction of the Voronoi tesselation at the half-time step.  Another is the follow the method of \citet{2016MNRAS.455.1134P} who adopted Heun's method to calculate the flux from the average of the fluxes at the beginning and end of the timestep.  The first method is straightforward to implement in \changaMM, but comes at the cost of another mesh-generation step.  The second method avoids the additional construction, and it may be as straightforward to implement (given the design of \changaMM), but this requires further investigation.}
  \item {\bf Generalized Equations of State:} Thus far, we have implemented an adiabatic equation of state $P\propto\rho^{\gamma}$, where $\gamma$ is the adiabatic index.  We have also adapted the Helmholtz equation of state \citep{2000ApJS..126..501T} to \changaMM, but this is still in testing.  In the near future, we aim to complete its implementation and also use an amalgamation of different equations of state suitable for studies in stellar physics similar to what is done in stellar evolution codes such as MESA \citep{2011ApJS..192....3P,2013ApJS..208....4P,2015ApJS..220...15P}.
  \item {\bf Individual Time Steps:} \changaMM\ as currently implemented imposes one universal time step.  We will leverage the individual time stepping that was developed for the SPH module in ChaNGa to allow individual timesteps for \changaMM.  We will use the same strategy as discussed in S10 and implement the individual timesteps in a pairwise fashion across each face which would allow the code to be far more efficient. We anticipate speedups between 2-4 for stellar mergers and possibly 1-2 orders of magnitude of improvement for other problems with high dynamic range.

  One issue is the determination of timesteps for each cell especially in the presence of large velocity fluctuations.  S10 proposed a scheme where the signal speed, the maximum speed by which hydrodynamic information can propagate, from each cell to every other cell is constructed in a tree-based manner (for efficiency reasons) to calculate the timestep.  Alternatively, \citet{2015MNRAS.450...53H} proposed calculating the signal speed over a local patch and incorporating the \citet{2009ApJ...697L..99S} criterion to smooth out timestepping changes over that local patch.  It is likely that we will choose this latter scheme to implement due to the greater simplicity by which it can be incorporated into ChaNGa.
  \item {\bf Cosmological Hydrodynamics:} As currently implemented, we assume that the scale factor $a$ is fixed at unity.  Modifications for a time variable scale factor would allow cosmological integrations.  We expect to include this physics as outlined by S10 in the near future.
  \item {\bf Physics for Isolated and Cosmological Galaxy Formation:} To target isolated and cosmological galaxy formation, we plan to either port or implement some additional physics.   This includes the \citet{2010MNRAS.407.1581S} prescription for metal line cooling, self-shielding \citep{2008MNRAS.390.1349P}, and turbulent diffusion of metals with a cosmic UV background \citep{1996ApJ...461...20H,2012ApJ...746..125H} and a subgrid model for star formation, thermal energy feedback, and blastwave supernovae feedback \citep{2006MNRAS.373.1074S}.
  \item {\bf SPH Initial Condition Translator:} As mentioned above, we currently use SPH initial conditions to initialize \changaMM.  These initial conditions are almost the same as regular SPH initial conditions with the noted exception that the SPH particles fill all of a well-defined periodic space.  However, this precludes the use of initial condition codes that already have been developed for ChaNGa. Toward that end, we plan on developing a translator that will fill in the atmosphere for SPH initial conditions in an adaptive manner similar to what AREPO does (S10).  
  \item {\bf Magnetic Fields:} S10's algorithm for unstructured MM hydrodynamics has been extend to magnetohydrodynamics by \citet{2011MNRAS.418.1392P}, \citet{2014MNRAS.442...43M}, and \citet{2016MNRAS.463..477M}.  
  We plan on implementing a suitable scheme in the near future.
  \item {\bf Improved Voronoi Tessellation Performance:} The Voronoi tessellation library, VORO++, used in \changaMM\ has very high performance, but its full performance capacity is not yet utilized as our test of completeness of the Voronoi cells are still relatively primitive.  Additional tests that incorporate the position of the point relative to the different vertices may help improve performance.  Further study on the effect of the domain decomposition on the construction of the Voronoi tessellation may also yield additional performance.

\end{enumerate}

Further down the road, we are interested in further developments of the algorithm including higher-order methods especially discontinuous Galerkin methods, and radiation (magneto-)hydrodynamics.

\section{Conclusions}\label{sec:conclusions}

We describe the structure and implementation of a MM hydrodynamics module, \changaMM, in the large-scale parallel code, ChaNGa.  Our implementation largely follows that of S10, with the notable exceptions of a direct Voronoi tessellation calculation, use of conserved rather than primitive variable for the reconstruction, a modified method of calculating the gradients that follows \citep{2016MNRAS.459.1596S}, reconstruct of the facial states using the half-time-step estimate of the state and a slope limiter that includes the \citet{2011ApJS..197...15D} slope limiter.  We have tested the code on a number of test problems and found that it produces consistent results with its SPH cousin, though in certain cases it appear to avoid some of the unphysical properties of (traditional) SPH.  We note, however, that more modern SPH solvers can avoid many of the pitfalls with traditional SPH \citep{Wadsley2017}.

The advent of MM codes that began with the development of AREPO (S10) has spurred tremendous interest in ALE codes in the last few years.  They seem ideally suited for astrophysics with seemingly superior treatment of instabilities and interfaces \citep[But also see][]{2015MNRAS.452.3853M} though recent work by \citet{2016MNRAS.455.4274L} suggest that the grid based codes, such as ATHENA \citep{2008ApJS..178..137S}, are excellent at sufficient resolution in this respect as well.  In addition, it can follow interesting regions efficiently over a large dynamical range.  It is hoped that the development of \changaMM\ which is planned to be released publicly and will include a full suite of different physics will spur further development and application of these MM codes into different areas of astrophysics.

\section*{Acknowledgments}

We thank Chris Rycroft for making VORO++ publicly available and for helpful and detailed discussions.  We thank Bert Vandenbroucke, R\"udiger Pakmor,  Volker Springel, and Elad Steinberg for useful discussions. PC thanks Paul Duffell for posting a tutorial on writing an Eulerian grid code which was extremely clear and helpful.  We thank the anonymous reviewer for comments and suggestions that improved this paper. 
PC is supported in part by the NASA ATP
program through NASA grant NNX13AH43G, NSF grant AST-1255469, and the University of Wisconsin-Milwaukee. JW acknowledges Natural Sciences and Engineering Research Council of Canada for funding support. TRQ was partially supported by NSF
award ACI-1550234.
Some of the computations were performed on
the gpc supercomputer at the SciNet HPC Consortium
\citep{2010JPhCS.256a2026L}. SciNet is funded by: the Canada
Foundation for Innovation under the auspices of Compute Canada; the
Government of Ontario; Ontario Research Fund - Research Excellence;
and the University of Toronto.
The authors also acknowledge the Texas Advanced Computing Center (TACC) at The University of Texas
at Austin for providing HPC resources that have contributed to the research results reported
within this paper. URL: \url{http://www.tacc.utexas.edu}



\bibliographystyle{mnras}
\bibliography{references}

\begin{thebibliography}{}
\makeatletter
\relax
\def\mn@urlcharsother{\let\do\@makeother \do\$\do\&\do\#\do\^\do\_\do\%\do\~}
\def\mn@doi{\begingroup\mn@urlcharsother \@ifnextchar [ {\mn@doi@}
  {\mn@doi@[]}}
\def\mn@doi@[#1]#2{\def\@tempa{#1}\ifx\@tempa\@empty \href
  {http://dx.doi.org/#2} {doi:#2}\else \href {http://dx.doi.org/#2} {#1}\fi
  \endgroup}
\def\mn@eprint#1#2{\mn@eprint@#1:#2::\@nil}
\def\mn@eprint@arXiv#1{\href {http://arxiv.org/abs/#1} {{\tt arXiv:#1}}}
\def\mn@eprint@dblp#1{\href {http://dblp.uni-trier.de/rec/bibtex/#1.xml}
  {dblp:#1}}
\def\mn@eprint@#1:#2:#3:#4\@nil{\def\@tempa {#1}\def\@tempb {#2}\def\@tempc
  {#3}\ifx \@tempc \@empty \let \@tempc \@tempb \let \@tempb \@tempa \fi \ifx
  \@tempb \@empty \def\@tempb {arXiv}\fi \@ifundefined
  {mn@eprint@\@tempb}{\@tempb:\@tempc}{\expandafter \expandafter \csname
  mn@eprint@\@tempb\endcsname \expandafter{\@tempc}}}

\bibitem[\protect\citeauthoryear{{Agertz} et~al.,}{{Agertz}
  et~al.}{2007}]{2007MNRAS.380..963A}
{Agertz} O.,  et~al., 2007, \mn@doi [\mnras]
  {10.1111/j.1365-2966.2007.12183.x}, \href
  {http://adsabs.harvard.edu/abs/2007MNRAS.380..963A} {380, 963}

\bibitem[\protect\citeauthoryear{{Balsara}}{{Balsara}}{1995}]{1995JCoPh.121..357B}
{Balsara} D.~S.,  1995, \mn@doi [Journal of Computational Physics]
  {10.1016/S0021-9991(95)90221-X}, \href
  {http://adsabs.harvard.edu/abs/1995JCoPh.121..357B} {121, 357}

\bibitem[\protect\citeauthoryear{{Barnes} \& {Hut}}{{Barnes} \&
  {Hut}}{1986}]{1986Natur.324..446B}
{Barnes} J.,  {Hut} P.,  1986, \mn@doi [\nat] {10.1038/324446a0}, \href
  {http://adsabs.harvard.edu/abs/1986Natur.324..446B} {324, 446}

\bibitem[\protect\citeauthoryear{{Borgers} \& {Peskin}}{{Borgers} \&
  {Peskin}}{1987}]{1987JCoPh..70..397B}
{Borgers} C.,  {Peskin} C.~S.,  1987, \mn@doi [Journal of Computational
  Physics] {10.1016/0021-9991(87)90189-6}, \href
  {http://adsabs.harvard.edu/abs/1987JCoPh..70..397B} {70, 397}

\bibitem[\protect\citeauthoryear{Donea, Huerta, Ponthot  \&
  Rodríguez-Ferran}{Donea et~al.}{2004}]{ALEBook}
Donea J.,  Huerta A.,  Ponthot J.-P.,   Rodríguez-Ferran A.,  2004, Arbitrary
  Lagrangian Eulerian Methods.
John Wiley \& Sons, Ltd, \mn@doi{10.1002/0470091355.ecm009}, \url
  {http://dx.doi.org/10.1002/0470091355.ecm009}

\bibitem[\protect\citeauthoryear{{Dubey} et~al.,}{{Dubey}
  et~al.}{2008}]{2008ASPC..385..145D}
{Dubey} A.,  et~al., 2008, in {Pogorelov} N.~V.,  {Audit} E.,   {Zank} G.~P.,
  eds,  Astronomical Society of the Pacific Conference Series Vol. 385,
  Numerical Modeling of Space Plasma Flows. p.~145

\bibitem[\protect\citeauthoryear{{Duffell}}{{Duffell}}{2016}]{2016ApJS..226....2D}
{Duffell} P.~C.,  2016, \mn@doi [\apjs] {10.3847/0067-0049/226/1/2}, \href
  {http://adsabs.harvard.edu/abs/2016ApJS..226....2D} {226, 2}

\bibitem[\protect\citeauthoryear{{Duffell} \& {MacFadyen}}{{Duffell} \&
  {MacFadyen}}{2011}]{2011ApJS..197...15D}
{Duffell} P.~C.,  {MacFadyen} A.~I.,  2011, \mn@doi [\apjs]
  {10.1088/0067-0049/197/2/15}, \href
  {http://adsabs.harvard.edu/abs/2011ApJS..197...15D} {197, 15}

\bibitem[\protect\citeauthoryear{{Fryxell} et~al.,}{{Fryxell}
  et~al.}{2000}]{2000ApJS..131..273F}
{Fryxell} B.,  et~al., 2000, \mn@doi [\apjs] {10.1086/317361}, \href
  {http://adsabs.harvard.edu/abs/2000ApJS..131..273F} {131, 273}

\bibitem[\protect\citeauthoryear{{Gaburov}, {Johansen}  \& {Levin}}{{Gaburov}
  et~al.}{2012}]{2012ApJ...758..103G}
{Gaburov} E.,  {Johansen} A.,   {Levin} Y.,  2012, \mn@doi [\apj]
  {10.1088/0004-637X/758/2/103}, \href
  {http://adsabs.harvard.edu/abs/2012ApJ...758..103G} {758, 103}

\bibitem[\protect\citeauthoryear{{Gnedin}}{{Gnedin}}{1995}]{1995ApJS...97..231G}
{Gnedin} N.~Y.,  1995, \mn@doi [\apjs] {10.1086/192141}, \href
  {http://adsabs.harvard.edu/abs/1995ApJS...97..231G} {97, 231}

\bibitem[\protect\citeauthoryear{{Gresho} \& {Chan}}{{Gresho} \&
  {Chan}}{1990}]{Gresho1990}
{Gresho} P.~M.,  {Chan} S.~T.,  1990, \mn@doi [International Journal for
  Numerical Methods in Fluids] {10.1002/fld.1650110510}, \href
  {http://adsabs.harvard.edu/abs/1990IJNMF..11..621G} {11, 621}

\bibitem[\protect\citeauthoryear{{Haardt} \& {Madau}}{{Haardt} \&
  {Madau}}{1996}]{1996ApJ...461...20H}
{Haardt} F.,  {Madau} P.,  1996, \mn@doi [\apj] {10.1086/177035}, \href
  {http://adsabs.harvard.edu/abs/1996ApJ...461...20H} {461, 20}

\bibitem[\protect\citeauthoryear{{Haardt} \& {Madau}}{{Haardt} \&
  {Madau}}{2012}]{2012ApJ...746..125H}
{Haardt} F.,  {Madau} P.,  2012, \mn@doi [\apj] {10.1088/0004-637X/746/2/125},
  \href {http://adsabs.harvard.edu/abs/2012ApJ...746..125H} {746, 125}

\bibitem[\protect\citeauthoryear{{Hopkins}}{{Hopkins}}{2013}]{2013MNRAS.428.2840H}
{Hopkins} P.~F.,  2013, \mn@doi [\mnras] {10.1093/mnras/sts210}, \href
  {http://adsabs.harvard.edu/abs/2013MNRAS.428.2840H} {428, 2840}

\bibitem[\protect\citeauthoryear{{Hopkins}}{{Hopkins}}{2015}]{2015MNRAS.450...53H}
{Hopkins} P.~F.,  2015, \mn@doi [\mnras] {10.1093/mnras/stv195}, \href
  {http://adsabs.harvard.edu/abs/2015MNRAS.450...53H} {450, 53}

\bibitem[\protect\citeauthoryear{Jetley, Gioachin, Mendes, Kale  \&
  Quinn}{Jetley et~al.}{2008}]{Jetley2008}
Jetley P.,  Gioachin F.,  Mendes C.,  Kale L.~V.,   Quinn T.~R.,  2008,
  Proceedings of IEEE International Parallel and Distributed Processing
  Symposium

\bibitem[\protect\citeauthoryear{Jetley, Wesolowski, Gioachin, Kale  \&
  Quinn}{Jetley et~al.}{2010}]{Jetley2010}
Jetley P.,  Wesolowski F.,  Gioachin F.,  Kale L.~V.,   Quinn T.~R.,  2010,
  Proceedings of the 2010 ACM/IEEE International Conference for High
  Performance Computin

\bibitem[\protect\citeauthoryear{Kale \& Krishnan}{Kale \&
  Krishnan}{1996}]{KaleKrishnan96}
Kale L.~V.,  Krishnan S.,  1996, in Wilson G.~V.,  Lu P.,  eds, , Parallel
  Programming using C++.
MIT Press, pp 175--213

\bibitem[\protect\citeauthoryear{{Kim} et~al.,}{{Kim}
  et~al.}{2014}]{2014ApJS..210...14K}
{Kim} J.-h.,  et~al., 2014, \mn@doi [\apjs] {10.1088/0067-0049/210/1/14}, \href
  {http://adsabs.harvard.edu/abs/2014ApJS..210...14K} {210, 14}

\bibitem[\protect\citeauthoryear{{Kim} et~al.,}{{Kim}
  et~al.}{2016}]{2016ApJ...833..202K}
{Kim} J.-h.,  et~al., 2016, \mn@doi [\apj] {10.3847/1538-4357/833/2/202}, \href
  {http://adsabs.harvard.edu/abs/2016ApJ...833..202K} {833, 202}

\bibitem[\protect\citeauthoryear{{Lecoanet} et~al.,}{{Lecoanet}
  et~al.}{2016}]{2016MNRAS.455.4274L}
{Lecoanet} D.,  et~al., 2016, \mn@doi [\mnras] {10.1093/mnras/stv2564}, \href
  {http://adsabs.harvard.edu/abs/2016MNRAS.455.4274L} {455, 4274}

\bibitem[\protect\citeauthoryear{Loken et~al.,}{Loken
  et~al.}{2010}]{2010JPhCS.256a2026L}
Loken C.,  et~al., 2010, Journal of Physics: Conference Series, 256, 012026

\bibitem[\protect\citeauthoryear{{Menon}, {Wesolowski}, {Zheng}, {Jetley},
  {Kale}, {Quinn}  \& {Governato}}{{Menon} et~al.}{2015}]{2015ComAC...2....1M}
{Menon} H.,  {Wesolowski} L.,  {Zheng} G.,  {Jetley} P.,  {Kale} L.,  {Quinn}
  T.,   {Governato} F.,  2015, \mn@doi [Computational Astrophysics and
  Cosmology] {10.1186/s40668-015-0007-9}, \href
  {http://adsabs.harvard.edu/abs/2015ComAC...2....1M} {2, 1}

\bibitem[\protect\citeauthoryear{{Mocz}, {Vogelsberger}  \& {Hernquist}}{{Mocz}
  et~al.}{2014}]{2014MNRAS.442...43M}
{Mocz} P.,  {Vogelsberger} M.,   {Hernquist} L.,  2014, \mn@doi [\mnras]
  {10.1093/mnras/stu865}, \href
  {http://adsabs.harvard.edu/abs/2014MNRAS.442...43M} {442, 43}

\bibitem[\protect\citeauthoryear{{Mocz}, {Vogelsberger}, {Pakmor}, {Genel},
  {Springel}  \& {Hernquist}}{{Mocz} et~al.}{2015}]{2015MNRAS.452.3853M}
{Mocz} P.,  {Vogelsberger} M.,  {Pakmor} R.,  {Genel} S.,  {Springel} V.,
  {Hernquist} L.,  2015, \mn@doi [\mnras] {10.1093/mnras/stv1598}, \href
  {http://adsabs.harvard.edu/abs/2015MNRAS.452.3853M} {452, 3853}

\bibitem[\protect\citeauthoryear{{Mocz}, {Pakmor}, {Springel}, {Vogelsberger},
  {Marinacci}  \& {Hernquist}}{{Mocz} et~al.}{2016}]{2016MNRAS.463..477M}
{Mocz} P.,  {Pakmor} R.,  {Springel} V.,  {Vogelsberger} M.,  {Marinacci} F.,
  {Hernquist} L.,  2016, \mn@doi [\mnras] {10.1093/mnras/stw2004}, \href
  {http://adsabs.harvard.edu/abs/2016MNRAS.463..477M} {463, 477}

\bibitem[\protect\citeauthoryear{{Monaghan}}{{Monaghan}}{1992}]{1992ARA&A..30..543M}
{Monaghan} J.~J.,  1992, \mn@doi [\araa] {10.1146/annurev.aa.30.090192.002551},
  \href {http://adsabs.harvard.edu/abs/1992ARA%26A..30..543M} {30, 543}

\bibitem[\protect\citeauthoryear{{Monaghan}}{{Monaghan}}{2005}]{2005RPPh...68.1703M}
{Monaghan} J.~J.,  2005, \mn@doi [Reports on Progress in Physics]
  {10.1088/0034-4885/68/8/R01}, \href
  {http://adsabs.harvard.edu/abs/2005RPPh...68.1703M} {68, 1703}

\bibitem[\protect\citeauthoryear{{Murphy} \& {Burrows}}{{Murphy} \&
  {Burrows}}{2008}]{2008ApJS..179..209M}
{Murphy} J.~W.,  {Burrows} A.,  2008, \mn@doi [\apjs] {10.1086/591272}, \href
  {http://adsabs.harvard.edu/abs/2008ApJS..179..209M} {179, 209}

\bibitem[\protect\citeauthoryear{Noh}{Noh}{1964}]{Noh1964}
Noh W.~F.,  1964, in Alder B.,  Fernbach S.,   Rotenberg M.,  eds, Methods in
  Computational Physics. Academic Press, pp 117--179

\bibitem[\protect\citeauthoryear{{Ohlmann}, {R{\"o}pke}, {Pakmor}  \&
  {Springel}}{{Ohlmann} et~al.}{2016}]{2016ApJ...816L...9O}
{Ohlmann} S.~T.,  {R{\"o}pke} F.~K.,  {Pakmor} R.,   {Springel} V.,  2016,
  \mn@doi [\apjl] {10.3847/2041-8205/816/1/L9}, \href
  {http://adsabs.harvard.edu/abs/2016ApJ...816L...9O} {816, L9}

\bibitem[\protect\citeauthoryear{{Pakmor}, {Bauer}  \& {Springel}}{{Pakmor}
  et~al.}{2011}]{2011MNRAS.418.1392P}
{Pakmor} R.,  {Bauer} A.,   {Springel} V.,  2011, \mn@doi [\mnras]
  {10.1111/j.1365-2966.2011.19591.x}, \href
  {http://adsabs.harvard.edu/abs/2011MNRAS.418.1392P} {418, 1392}

\bibitem[\protect\citeauthoryear{{Pakmor}, {Springel}, {Bauer}, {Mocz},
  {Munoz}, {Ohlmann}, {Schaal}  \& {Zhu}}{{Pakmor}
  et~al.}{2016a}]{2016MNRAS.455.1134P}
{Pakmor} R.,  {Springel} V.,  {Bauer} A.,  {Mocz} P.,  {Munoz} D.~J.,
  {Ohlmann} S.~T.,  {Schaal} K.,   {Zhu} C.,  2016a, \mn@doi [\mnras]
  {10.1093/mnras/stv2380}, \href
  {http://adsabs.harvard.edu/abs/2016MNRAS.455.1134P} {455, 1134}

\bibitem[\protect\citeauthoryear{{Pakmor}, {Pfrommer}, {Simpson}, {Kannan}  \&
  {Springel}}{{Pakmor} et~al.}{2016b}]{2016MNRAS.462.2603P}
{Pakmor} R.,  {Pfrommer} C.,  {Simpson} C.~M.,  {Kannan} R.,   {Springel} V.,
  2016b, \mn@doi [\mnras] {10.1093/mnras/stw1761}, \href
  {http://adsabs.harvard.edu/abs/2016MNRAS.462.2603P} {462, 2603}

\bibitem[\protect\citeauthoryear{{Paxton}, {Bildsten}, {Dotter}, {Herwig},
  {Lesaffre}  \& {Timmes}}{{Paxton} et~al.}{2011}]{2011ApJS..192....3P}
{Paxton} B.,  {Bildsten} L.,  {Dotter} A.,  {Herwig} F.,  {Lesaffre} P.,
  {Timmes} F.,  2011, \mn@doi [\apjs] {10.1088/0067-0049/192/1/3}, \href
  {http://adsabs.harvard.edu/abs/2011ApJS..192....3P} {192, 3}

\bibitem[\protect\citeauthoryear{{Paxton} et~al.,}{{Paxton}
  et~al.}{2013}]{2013ApJS..208....4P}
{Paxton} B.,  et~al., 2013, \mn@doi [\apjs] {10.1088/0067-0049/208/1/4}, \href
  {http://adsabs.harvard.edu/abs/2013ApJS..208....4P} {208, 4}

\bibitem[\protect\citeauthoryear{{Paxton} et~al.,}{{Paxton}
  et~al.}{2015}]{2015ApJS..220...15P}
{Paxton} B.,  et~al., 2015, \mn@doi [\apjs] {10.1088/0067-0049/220/1/15}, \href
  {http://adsabs.harvard.edu/abs/2015ApJS..220...15P} {220, 15}

\bibitem[\protect\citeauthoryear{{Pen}}{{Pen}}{1998}]{1998ApJS..115...19P}
{Pen} U.-L.,  1998, \mn@doi [\apjs] {10.1086/313074}, \href
  {http://adsabs.harvard.edu/abs/1998ApJS..115...19P} {115, 19}

\bibitem[\protect\citeauthoryear{{Pfrommer}, {Pakmor}, {Schaal}, {Simpson}  \&
  {Springel}}{{Pfrommer} et~al.}{2017}]{2017MNRAS.465.4500P}
{Pfrommer} C.,  {Pakmor} R.,  {Schaal} K.,  {Simpson} C.~M.,   {Springel} V.,
  2017, \mn@doi [\mnras] {10.1093/mnras/stw2941}, \href
  {http://adsabs.harvard.edu/abs/2017MNRAS.465.4500P} {465, 4500}

\bibitem[\protect\citeauthoryear{{Pontzen} et~al.,}{{Pontzen}
  et~al.}{2008}]{2008MNRAS.390.1349P}
{Pontzen} A.,  et~al., 2008, \mn@doi [\mnras]
  {10.1111/j.1365-2966.2008.13782.x}, \href
  {http://adsabs.harvard.edu/abs/2008MNRAS.390.1349P} {390, 1349}

\bibitem[\protect\citeauthoryear{{Rycroft}}{{Rycroft}}{2009}]{2009Chaos..19d1111R}
{Rycroft} C.~H.,  2009, \mn@doi [Chaos] {10.1063/1.3215722}, \href
  {http://adsabs.harvard.edu/abs/2009Chaos..19d1111R} {19, 041111}

\bibitem[\protect\citeauthoryear{{Saitoh} \& {Makino}}{{Saitoh} \&
  {Makino}}{2009}]{2009ApJ...697L..99S}
{Saitoh} T.~R.,  {Makino} J.,  2009, \mn@doi [\apjl]
  {10.1088/0004-637X/697/2/L99}, \href
  {http://adsabs.harvard.edu/abs/2009ApJ...697L..99S} {697, L99}

\bibitem[\protect\citeauthoryear{{Saitoh} \& {Makino}}{{Saitoh} \&
  {Makino}}{2013}]{2013ApJ...768...44S}
{Saitoh} T.~R.,  {Makino} J.,  2013, \mn@doi [\apj]
  {10.1088/0004-637X/768/1/44}, \href
  {http://adsabs.harvard.edu/abs/2013ApJ...768...44S} {768, 44}

\bibitem[\protect\citeauthoryear{{Shen}, {Wadsley}  \& {Stinson}}{{Shen}
  et~al.}{2010}]{2010MNRAS.407.1581S}
{Shen} S.,  {Wadsley} J.,   {Stinson} G.,  2010, \mn@doi [\mnras]
  {10.1111/j.1365-2966.2010.17047.x}, \href
  {http://adsabs.harvard.edu/abs/2010MNRAS.407.1581S} {407, 1581}

\bibitem[\protect\citeauthoryear{{Springel}}{{Springel}}{2005}]{2005MNRAS.364.1105S}
{Springel} V.,  2005, \mn@doi [\mnras] {10.1111/j.1365-2966.2005.09655.x},
  \href {http://adsabs.harvard.edu/abs/2005MNRAS.364.1105S} {364, 1105}

\bibitem[\protect\citeauthoryear{{Springel}}{{Springel}}{2010}]{2010MNRAS.401..791S}
{Springel} V.,  2010, \mn@doi [\mnras] {10.1111/j.1365-2966.2009.15715.x},
  \href {http://adsabs.harvard.edu/abs/2010MNRAS.401..791S} {401, 791}

\bibitem[\protect\citeauthoryear{{Stadel}}{{Stadel}}{2001}]{2001PhDT........21S}
{Stadel} J.~G.,  2001, PhD thesis, UNIVERSITY OF WASHINGTON

\bibitem[\protect\citeauthoryear{{Steinberg}, {Yalinewich}  \&
  {Sari}}{{Steinberg} et~al.}{2016}]{2016MNRAS.459.1596S}
{Steinberg} E.,  {Yalinewich} A.,   {Sari} R.,  2016, \mn@doi [\mnras]
  {10.1093/mnras/stw783}, \href
  {http://adsabs.harvard.edu/abs/2016MNRAS.459.1596S} {459, 1596}

\bibitem[\protect\citeauthoryear{{Stinson}, {Seth}, {Katz}, {Wadsley},
  {Governato}  \& {Quinn}}{{Stinson} et~al.}{2006}]{2006MNRAS.373.1074S}
{Stinson} G.,  {Seth} A.,  {Katz} N.,  {Wadsley} J.,  {Governato} F.,   {Quinn}
  T.,  2006, \mn@doi [\mnras] {10.1111/j.1365-2966.2006.11097.x}, \href
  {http://adsabs.harvard.edu/abs/2006MNRAS.373.1074S} {373, 1074}

\bibitem[\protect\citeauthoryear{{Stone}, {Gardiner}, {Teuben}, {Hawley}  \&
  {Simon}}{{Stone} et~al.}{2008}]{2008ApJS..178..137S}
{Stone} J.~M.,  {Gardiner} T.~A.,  {Teuben} P.,  {Hawley} J.~F.,   {Simon}
  J.~B.,  2008, \mn@doi [\apjs] {10.1086/588755}, \href
  {http://adsabs.harvard.edu/abs/2008ApJS..178..137S} {178, 137}

\bibitem[\protect\citeauthoryear{{Teyssier}}{{Teyssier}}{2002}]{2002A&A...385..337T}
{Teyssier} R.,  2002, \mn@doi [\aap] {10.1051/0004-6361:20011817}, \href
  {http://adsabs.harvard.edu/abs/2002A%26A...385..337T} {385, 337}

\bibitem[\protect\citeauthoryear{{Timmes} \& {Swesty}}{{Timmes} \&
  {Swesty}}{2000}]{2000ApJS..126..501T}
{Timmes} F.~X.,  {Swesty} F.~D.,  2000, \mn@doi [\apjs] {10.1086/313304}, \href
  {http://adsabs.harvard.edu/abs/2000ApJS..126..501T} {126, 501}

\bibitem[\protect\citeauthoryear{Toro}{Toro}{2009}]{toro2009riemann}
Toro E.,  2009, Riemann Solvers and Numerical Methods for Fluid Dynamics: A
  Practical Introduction.
Springer Berlin Heidelberg

\bibitem[\protect\citeauthoryear{{Turk}, {Smith}, {Oishi}, {Skory}, {Skillman},
  {Abel}  \& {Norman}}{{Turk} et~al.}{2011}]{2011ApJS..192....9T}
{Turk} M.~J.,  {Smith} B.~D.,  {Oishi} J.~S.,  {Skory} S.,  {Skillman} S.~W.,
  {Abel} T.,   {Norman} M.~L.,  2011, \mn@doi [The Astrophysical Journal
  Supplement Series] {10.1088/0067-0049/192/1/9}, \href
  {http://adsabs.harvard.edu/abs/2011ApJS..192....9T} {192, 9}

\bibitem[\protect\citeauthoryear{{Vandenbroucke} \& {De
  Rijcke}}{{Vandenbroucke} \& {De Rijcke}}{2016}]{2016A&C....16..109V}
{Vandenbroucke} B.,  {De Rijcke} S.,  2016, \mn@doi [Astronomy and Computing]
  {10.1016/j.ascom.2016.05.001}, \href
  {http://adsabs.harvard.edu/abs/2016A%26C....16..109V} {16, 109}

\bibitem[\protect\citeauthoryear{{Vogelsberger} et~al.,}{{Vogelsberger}
  et~al.}{2014}]{2014MNRAS.444.1518V}
{Vogelsberger} M.,  et~al., 2014, \mn@doi [\mnras] {10.1093/mnras/stu1536},
  \href {http://adsabs.harvard.edu/abs/2014MNRAS.444.1518V} {444, 1518}

\bibitem[\protect\citeauthoryear{{Wadsley}, {Stadel}  \& {Quinn}}{{Wadsley}
  et~al.}{2004}]{2004NewA....9..137W}
{Wadsley} J.~W.,  {Stadel} J.,   {Quinn} T.,  2004, \mn@doi [\na]
  {10.1016/j.newast.2003.08.004}, \href
  {http://adsabs.harvard.edu/abs/2004NewA....9..137W} {9, 137}

\bibitem[\protect\citeauthoryear{{Wadsley}, {Keller}  \& {Quinn}}{{Wadsley}
  et~al.}{2017}]{Wadsley2017}
{Wadsley} J.~W.,  {Keller} B.~W.,   {Quinn} T.~R.,  2017, preprint, \href
  {http://adsabs.harvard.edu/abs/2017arXiv170703824W} {} (\mn@eprint {arXiv}
  {1707.03824})

\bibitem[\protect\citeauthoryear{{Whitehurst}}{{Whitehurst}}{1995}]{1995MNRAS.277..655W}
{Whitehurst} R.,  1995, \mn@doi [\mnras] {10.1093/mnras/277.2.655}, \href
  {http://adsabs.harvard.edu/abs/1995MNRAS.277..655W} {277, 655}

\bibitem[\protect\citeauthoryear{{Yalinewich}, {Steinberg}  \&
  {Sari}}{{Yalinewich} et~al.}{2015}]{2015ApJS..216...35Y}
{Yalinewich} A.,  {Steinberg} E.,   {Sari} R.,  2015, \mn@doi [\apjs]
  {10.1088/0067-0049/216/2/35}, \href
  {http://adsabs.harvard.edu/abs/2015ApJS..216...35Y} {216, 35}

\bibitem[\protect\citeauthoryear{{Zhu}, {Pakmor}, {van Kerkwijk}  \&
  {Chang}}{{Zhu} et~al.}{2015}]{2015ApJ...806L...1Z}
{Zhu} C.,  {Pakmor} R.,  {van Kerkwijk} M.~H.,   {Chang} P.,  2015, \mn@doi
  [\apjl] {10.1088/2041-8205/806/1/L1}, \href
  {http://adsabs.harvard.edu/abs/2015ApJ...806L...1Z} {806, L1}

\makeatother
\end{thebibliography}

\bsp    
\label{lastpage}

\end{document}